\documentclass[aps,prmaterials,preprint,superscriptaddress]{revtex4-2}
\usepackage{gensymb}
\usepackage{graphicx}
\usepackage{booktabs}
\usepackage{multirow}
\usepackage{chemformula}
\usepackage{appendix}
\usepackage{xcolor}
\begin{document}
\title{Bulk magnetic properties of distorted square lattice compounds $M'$-$Ln\mathrm{TaO}_4$ ($Ln=$ Tb, Dy, Ho, Er)}
\author{Nicola D.~Kelly}
\email[]{ne281@cam.ac.uk}
\affiliation{Cavendish Laboratory, University of Cambridge, J J Thomson Avenue, Cambridge, CB3 0HE, United Kingdom}

\author{Ivan da Silva}
\affiliation{ISIS Neutron and Muon Source, Rutherford Appleton Laboratory, Didcot, OX11 0QX, United Kingdom}

\author{Si\^{a}n E.~Dutton}
\email[]{sed33@cam.ac.uk}
\affiliation{Cavendish Laboratory, University of Cambridge, J J Thomson Avenue, Cambridge, CB3 0HE, United Kingdom}

\date{\today}

\begin{abstract}
We report bulk magnetic properties of the monoclinic lanthanide tantalates, $M'$-\ch{\textit{Ln}TaO4} ($Ln$= Tb, Dy, Ho, Er), where the magnetic $Ln^{3+}$ ions are arranged on a distorted 2D square lattice. The heavier analogue $M'$-\ch{YbTaO4} has been investigated as a spin-orbit-coupled, quasi-two-dimensional frustrated magnet, and the properties of the other $M'$-\ch{\textit{Ln}TaO4} are expected to vary depending on the electronic configuration of the $Ln$ ion, namely Kramers vs non-Kramers behaviour and different crystal electric field parameters. In this work, powder neutron diffraction is used to confirm the crystal structure for $Ln=$ Tb, Ho, Er, and to determine the magnetic structure of $M'$-\ch{TbTaO4}, which displays long-range antiferromagnetic (AFM) order below $T_\mathrm{N}=2.1$~K. The Tb$^{3+}$ moments are aligned primarily along the $c$-axis with AFM nearest-neighbour interactions. Susceptibility data suggest that $M'$-\ch{DyTaO4} may display short-range ordering around 2.7~K, while $M'$-\ch{HoTaO4} and $M'$-\ch{ErTaO4} show AFM correlations but do not order above 1.8~K. Measurements of the magnetic specific heat provide evidence for a Kramers doublet ground states in $M'$-\ch{ErTaO4}, similar to its heavier analogue $M'$-\ch{YbTaO4}.

\end{abstract}
\maketitle

\section{Introduction}

The spin-$\frac{1}{2}$ Heisenberg square lattice is an important model for frustrated magnetism studies. The two relevant interactions are the nearest-neighbour coupling $J_1$, along the sides of the squares, and next-nearest-neighbour $J_2$, along the diagonals of the squares. Depending on the relative signs and magnitudes of $J_1$ and $J_2$, three types of classical long-range order (LRO) can be observed (ferromagnetic (FM), columnar antiferromagnetic (AFM), and N\'{e}el AFM), and there are also two boundary regions where spin-liquid behaviour is predicted 
\cite{Anderson1987,Greedan2001,Chamorro2021}. These arise from the competition between $J_1$ and $J_2$ when $J_2/J_1$ is close to $\pm 0.5$. Several experimental realisations of the Heisenberg square lattice have been studied recently, including \ch{YbBi2IO4} and \ch{YbBi2ClO4} \cite{Park2024}, \ch{SrLaCuSbO6} and \ch{SrLaCuNbO6} \cite{Watanabe2022}, and $\alpha$-\ch{KTi(C2O4)2.\textit{x}H2O} \cite{Abdeldaim2020}, which all order antiferromagnetically at low temperatures. However, a small number of materials have been found to resist order and are therefore considered candidates for the spin-liquid behaviour predicted at $J_2/J_1\approx\pm\frac{1}{2}$, including transition-metal oxides \ch{Sr2Cu(Te$_{1-x}$W$_x$)O6} \cite{Mustonen2018,Mustonen2018a} and a rare-earth oxide \ch{NdKNaNbO5} \cite{Guchhait2024}.

\begin{figure}[htbp]
\centering
\includegraphics[width=0.5\textwidth]{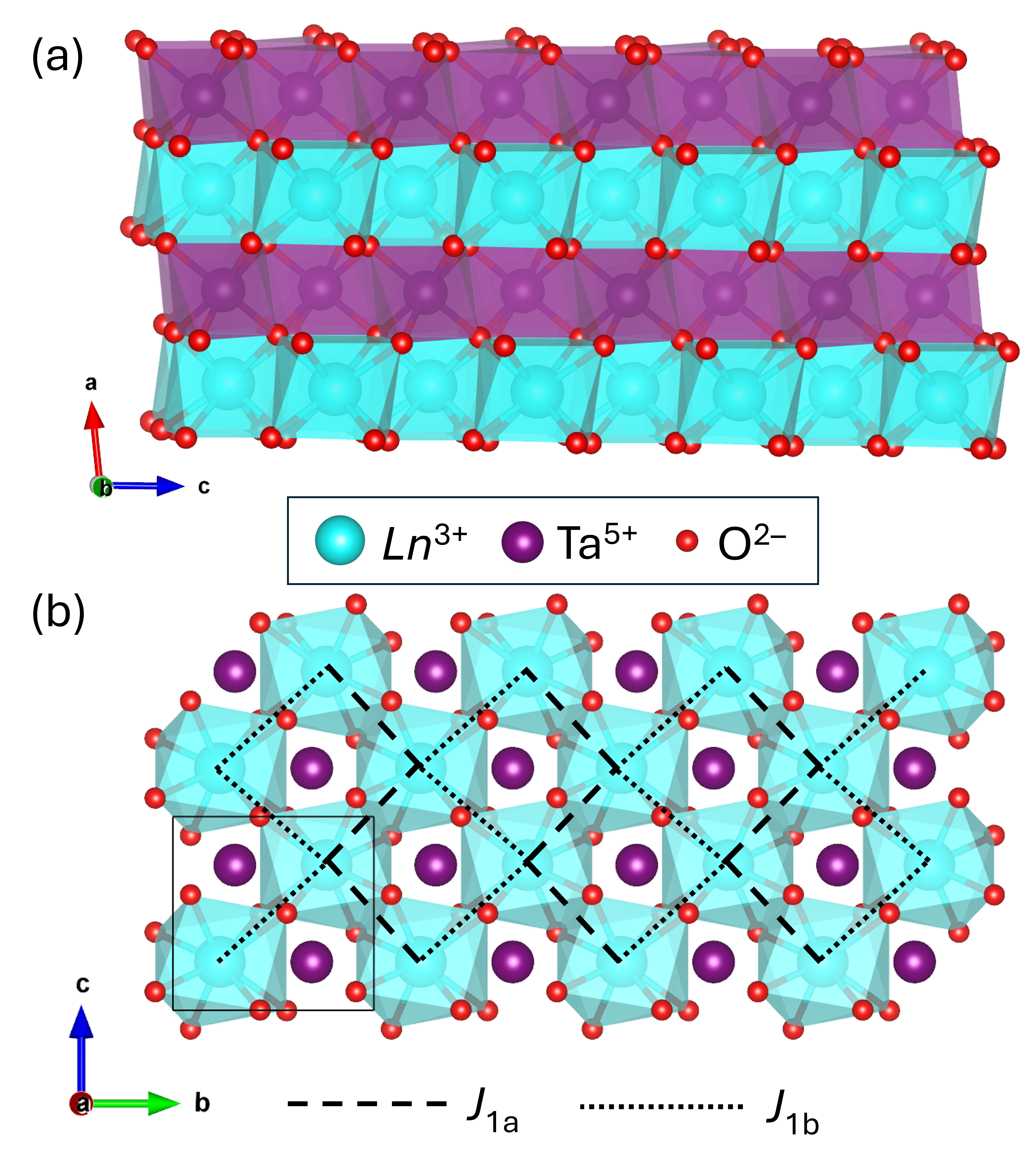}
\caption{Crystal structure of $M'$-\ch{\textit{Ln}TaO4}, space group $P2/c$ (No.~13), viewed along different crystallographic axes ($b$ and $a$ respectively). Blue atoms are $Ln^{3+}$, purple are \ch{Ta^{5+}} and red are \ch{O^{2-}}. In (b), dashed and dotted lines highlight the distorted square lattice of $Ln^{3+}$ ions.}
\label{fig:crystalstructure}
\end{figure}

This work focuses on the lanthanide orthotantalates, \ch{\textit{Ln}TaO4}, with the monoclinic $M'$ crystal structure. The tantalates crystallise in several different structures depending on the lanthanide ion, temperature and/or pressure. Compounds of the largest lanthanides, La--Pr, have the monoclinic space group $P2_1/c$, while compounds of the later (smaller) lanthanides can form the $M$ ($I2/a$) or $M'$ ($P2/c$) monoclinic phases or the tetragonal $T$ ($I4_1/a$) phase (which is only observed \textit{in situ}) \cite{Brixner1983}. The magnetic $Ln^{3+}$ ions in $M$-\ch{\textit{Ln}TaO4} occupy a three-dimensional (3D) distorted diamond lattice, which leads to geometric magnetic frustration from the competing nearest-neighbour interactions \cite{Kelly2022a,Zhang2025}. In contrast, the layered $M'$ phase contains magnetic $Ln^{3+}$ ions on a two-dimensional (2D) distorted square lattice, which also displays frustration.

The crystal structure of $M'$-\ch{\textit{Ln}TaO4} is shown in Fig.~\ref{fig:crystalstructure}. It is built from alternating layers of edge-sharing \ch{TaO6} octahedra and \ch{\textit{Ln}O8} square antiprisms, Fig.~\ref{fig:crystalstructure}(a). Within each layer of lanthanide-oxygen polyhedra, the $Ln^{3+}$ ions form a distorted square network, Fig.~\ref{fig:crystalstructure}(b), with two distinct nearest-neighbour distances, denoted $J_{1a}$ and $J_{1b}$. The interlayer $Ln$--$Ln$ separation (e.g.~5.15~\AA\ for $M'$-\ch{TbTaO4} at room temperature) is more than 30~\%\ longer than either of the intralayer $Ln$--$Ln$ distances (3.71 and 3.96~\AA\ for $M'$-\ch{TbTaO4}). Therefore, we can consider this a quasi-2D system. The combination of highly localised $4f$ orbitals, a quasi-2D magnetic network, and geometric frustration makes it likely that long-range order might be partially or wholly suppressed in these materials.

Recently, two independent studies investigated frustrated magnetism in the isostructural spin-$\frac{1}{2}$ system, $M'$-\ch{YbTaO4}. They both determined that \ch{YbTaO4} has a spin-orbit-driven $J_{eff}=\frac{1}{2}$ Kramers doublet ground state with a large gap to the first excited state. The Curie-Weiss temperature $\theta_{CW}$ was negative, indicating weak antiferromagnetic correlations between \ch{Yb^{3+}} spins. Ramanathan \textit{et al}.\ found no magnetic ordering down to 100~mK in specific heat measurements \cite{Ramanathan2024}. This means that the frustration index $f=|\theta_{CW}/T_\mathrm{N}|\approx 10$, indicating strong frustration \cite{Ramirez1994}. Kumar \textit{et al}.\ similarly found a lack of magnetic order at $T\geq1.8$~K; they also confirmed the presence of fluctuating \ch{Yb^{3+}} spins using muon-spin relaxation ($\mu$SR) spectroscopy \cite{Kumar2024}. The combined experimental evidence supports a quantum disordered ground state in $M'$-\ch{YbTaO4}.

In this study, we measure the bulk magnetic properties of the analogous lanthanide tantalate compounds with $Ln=$ Tb, Dy, Ho and Er, which each have unique electronic configurations and different quantum numbers $L$, $S$ and $J$. We find that $M'$-\ch{TbTaO4} orders antiferromagnetically at 2.1~K, while the other three compounds have AFM correlations (Curie-Weiss constants ranging from $-3$ to $-7.5$~K) but do not display long-range order above 1.8~K. Neutron diffraction was used to confirm the nuclear crystal structures at room temperature for $Ln=$ Ho and Er. Neutron diffraction also allowed us to track the nuclear structure of $M'$-\ch{TbTaO4} as a function of temperature on cooling to 1.5~K, and to determine its magnetic structure below $T_\mathrm{N}$. We find that $M'$-\ch{TbTaO4} displays long-range AFM order with an ordering vector of $\vec{k}=(\frac{1}{2},\frac{1}{2},0)$ and the spins directed mainly along the $c$-direction.

\section{Experimental Methods}

\subsection{Synthesis}
Polycrystalline samples of \ch{\textit{Ln}TaO4} in the monoclinic $M'$ phase were synthesised using ceramic methods on a 0.5--4.0~g scale. The lanthanide oxides, \ch{Tb4O7} (99.9~\%, Alfa Aesar), \ch{Dy2O3} (99.99~\%, Alfa Aesar), \ch{Ho2O3} (99.99~\%, Alfa Aesar) or \ch{Er2O3} (99.9~\%, Alfa Aesar) respectively, were ground in stoichiometric ratio with \ch{Ta2O5} (99.993~\%, Thermo Scientific) using a pestle and mortar. The powders were placed in alumina crucibles and heated in air at 3~\degree C min$^{-1}$ to 1200--1450~\degree C for 24 h (a more detailed discussion of the optimum temperature for each sample can be found in Section~\ref{section:structure}). The samples were cooled to room temperature at the natural rate of the furnace, ground and reheated with the same heating profile until the weight fractions of impurity phases no longer decreased.

\subsection{Powder X-ray diffraction (PXRD)}
Phase purity was checked using powder X-ray diffraction (PXRD) on a Bruker D8 diffractometer with Cu K$\alpha$ radiation, $\lambda = 1.541$~\AA. PXRD patterns for structural refinement were collected at the I11 beamline, Diamond Light Source \cite{Thompson2009a}, with wavelength (refined from a silicon standard) $\lambda= 0.82869$~\AA. Samples were mixed 1:1 by volume with ground glass to reduce absorption and packed into $\phi=0.5$~mm borosilicate capillaries. Rietveld refinement \cite{Rietveld1969} was carried out using the program TOPAS v5 \cite{Coelho2018}. The background was fitted with a Chebyshev polynomial with 12 coefficients and the peakshape was a Voigt function with axial divergence asymmetry. During refinement, the $x$ and $z$ coordinates of the $Ln$ and Ta atoms were fixed at these values representing special positions in the cell: $x_{Ln}=0$, $z_{Ln}=0.25$; $x_\mathrm{Ta}=0.5$, $z_\mathrm{Ta}=0.75$. The atomic coordinates of the oxygen atoms were also fixed at values taken from the literature \cite{Hartenbach2005}. The refined parameters included the background coefficients, zero offset, peak shape parameters, $y$ coordinates of the $Ln$ and Ta atoms, and isotropic thermal parameters $B_\mathrm{iso}$ for each element.

\subsection{Powder neutron diffraction (PND)}
Large samples (approximately 4~g each) of $M'$-\ch{TbTaO4}, $M'$-\ch{HoTaO4} and $M'$-\ch{ErTaO4} were investigated using time-of-flight (TOF) powder neutron diffraction (PND) at the GEM beamline \cite{Hannon2005}, ISIS Neutron and Muon Source \cite{GEMdata}. Note that $M'$-\ch{DyTaO4} was excluded from the neutron diffraction study because of the very strong neutron absorption of some Dy isotopes \cite{Sears1992} which would require isotopically enriched samples for PND experiments. The samples were packed into $\phi = 6$~mm vanadium cans and measured at 300~K (all samples) and at 100, 30 and 1.5~K (Tb only) using an ILL Orange cryostat \cite{Orange}. Rietveld refinement \cite{Rietveld1969} was carried out using the program TOPAS v5 \cite{Coelho2018} against banks 3--6 simultaneously, each with equal weighting.

\subsection{Magnetometry}
Magnetic measurements were made using a Quantum Design Materials Properties Measurement System (MPMS-3). Approximately 10~mg of each sample (weighed accurately in each case) was contained in clingfilm and the standard plastic sample holders then inserted into a brass sample holder. The DC magnetic moment was measured as a function of temperature upon warming in the range 1.8--300~K at a field of 500~Oe, after cooling in zero field (ZFC) or in the applied field (FC). The DC moment was also measured as a function of magnetic field in the range 0--7~T at temperatures of 2, 4, 6, 8, 10 and 100~K.

\subsection{Specific heat}
Approximately 40~mg of each sample (weighed accurately in each case) was ground with an equal mass of Ag powder (Alfa Aesar, 99.99\%, --635 mesh) using a pestle and mortar, then pressed into a $\phi=5$~mm pellet. Portions of this pellet weighing 5--20~mg were attached to the sample holder using Apiezon N grease. The heat capacity was measured at $T=$ 1.8--30~K using a Quantum Design Physical Properties Measurement System (PPMS); an initial addenda (background) measurement was made before adding the pellet, in order to subtract the heat capacity contribution arising from the grease. The Ag lattice contribution to the total heat capacity was then subtracted using values from the literature \cite{Smith1995}. Furthermore, for each compound $M'$-\ch{\textit{Ln}TaO4}, the lattice contribution $C_\mathrm{latt}$ was modelled using the Debye law:

\begin{equation}
C_\mathrm{latt}=\frac{9nRT^3}{{\theta_\mathrm{D}^3}}\int_{0}^{\frac{\theta_\mathrm{D}}{T}}\frac{x^4e^x}{(e^x-1)^2} dx
\end{equation}
where $n$ is the number of atoms per formula unit, $R$ is the molar gas constant, $T$ the temperature in K, and $\theta_\mathrm{D}$ the Debye temperature in K \cite{Gopal1966}. A scalar parameter $p$ was substituted for $9nR$ in the fitting equation and refined to account for the sample mass error and imperfect coupling to the sample stage. The lattice contribution was then subtracted from all datasets to leave the magnetic heat capacity, $C_\mathrm{mag}(T)$, for each compound. The obtained Debye temperatures were 256(2), 197(2), 127(3), and 148(2)~K for $Ln=$ Tb, Dy, Ho, and Er respectively.

\section{Results and Discussion}

\subsection{\label{section:structure}Structural characterisation}
The phase space of the lanthanide tantalates \ch{\textit{Ln}TaO4} contains several distinct polymorphs. For $Ln=$ Tb--Er, the equilibrium phase at room temperature is $M'$, which on heating to around 1400~\degree C transforms into the tetragonal $T$ (scheelite) phase. However, the $T$ phase cannot be isolated at room temperature and upon cooling transforms into the $M$ (fergusonite) phase \cite{Brixner1983,Mather1995}. Therefore, targeted synthesis of the $M'$ phase requires the temperature to be maintained as high as possible, to facilitate atomic diffusion, whilst not exceeding the $M'-T$ transition temperature \cite{Mullens2023,Mullens2024}. The optimum furnace temperature was found to be 1350~\degree C for $Ln=$ Tb, Dy and Ho, and 1450~\degree C for $Ln=$ Er. It proved impossible to stabilise phase-pure $M'$-\ch{GdTaO4} (or any pure $M'$ samples with larger lanthanide ions) because when the synthesis temperature was kept low enough to avoid forming the $M$ phase, there was always competition from \ch{Gd3TaO7} \cite{Wakeshima2004}.

The crystal structures of $M'$-\ch{\textit{Ln}TaO4} were refined against high-resolution synchrotron powder X-ray diffraction (PXRD) data using the Rietveld method. The refined parameters for all four compounds are given in Table~\ref{table:mlntao4_params} and the refinement plots are in the Supplemental Material, Figures S1--S4 \cite{Supplemental}. The PXRD data showed no presence of the competing $M$-phase in these samples and the refined structural parameters of the four samples agree well with reports from the literature \cite{Brixner1983,Hartenbach2005}. Note: with respect to those literature reports, the origin has been shifted by half a unit cell in the $b$ direction in order to achieve consistency between the nuclear and magnetic unit cell models (see later Section~\ref{section:magpnd} and Table S2 in Supplemental Material \cite{Supplemental}). 

\begin{table}[htbp]
\centering
\caption{Unit cell parameters for $M'$-\ch{\textit{Ln}TaO4} ($Ln=$ Tb, Dy, Ho, Er) in space group $P2/c$. Values obtained by Rietveld refinement against synchrotron PXRD data, $T=300$~K, $\lambda=0.82869$~\AA.}
\label{table:mlntao4_params}
\resizebox{\columnwidth}{!}{
\begin{ruledtabular}
\begin{tabular}{c c c c c}
 & $M'$-\ch{TbTaO4} & $M'$-\ch{DyTaO4} & $M'$-\ch{HoTaO4} & $M'$-\ch{ErTaO4} \\
\midrule
$a$ (\AA) & 5.14126(4) & 5.12917(2) & 5.11433(2) & 5.09892(4) \\
$b$ (\AA) & 5.48935(4) & 5.47022(3) & 5.45273(2) & 5.43875(5) \\
$c$ (\AA) & 5.33646(4) & 5.31716(2) & 5.30149(2) & 5.28448(5) \\
$\beta$ (\degree) & 96.6668(5) & 96.5857(3) & 96.5011(2) & 96.4289(4) \\
$V$ (\AA$^3$) & 149.588(2) & 148.203(1) & 146.893(1) & 145.626(2) \\
$y_{Ln}$ & 0.73406(16) & 0.73436(15) & 0.73385(13) & 0.73369(18) \\
$y_\mathrm{Ta}$ & 0.80846(15) & 0.80704(14) & 0.80600(12) & 0.80578(17) \\
$\chi^2$ & 3.68 & 3.98 & 4.72 & 4.21 \\
$R_\mathrm{wp}$ (\%) & 2.65 & 2.93 & 3.26 & 2.88 \\
\end{tabular}
\end{ruledtabular}
}
\end{table}

\begin{figure}[htbp]
\centering
\includegraphics[width=0.5\textwidth]{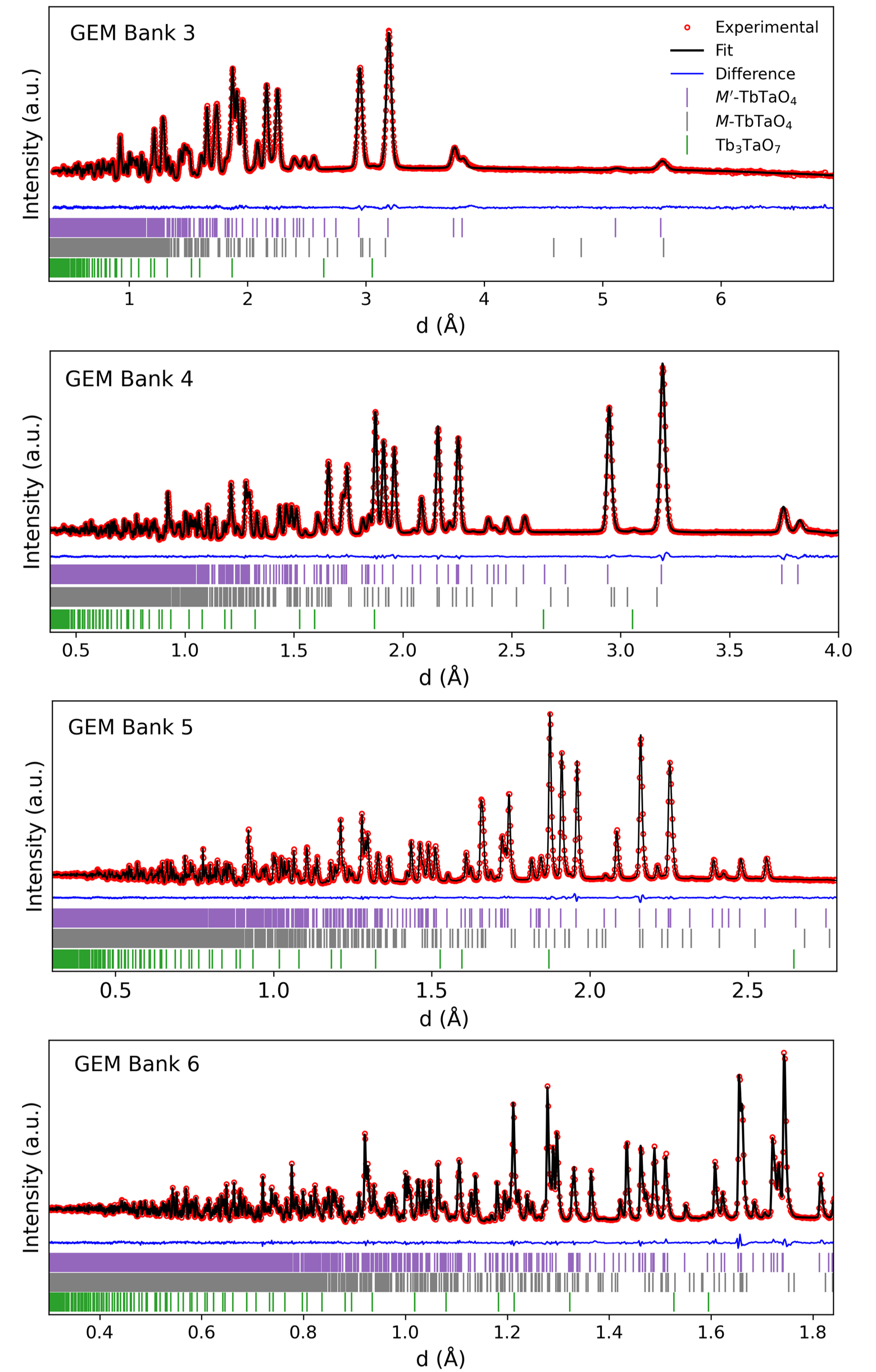}
\caption{Rietveld refinement against room-temperature TOF PND data for $M'$-\ch{TbTaO4}.}
\label{fig:rietveld_tb_pnd_rt}
\end{figure}

\begin{figure}[htbp]
\centering
\includegraphics[width=0.5\textwidth]{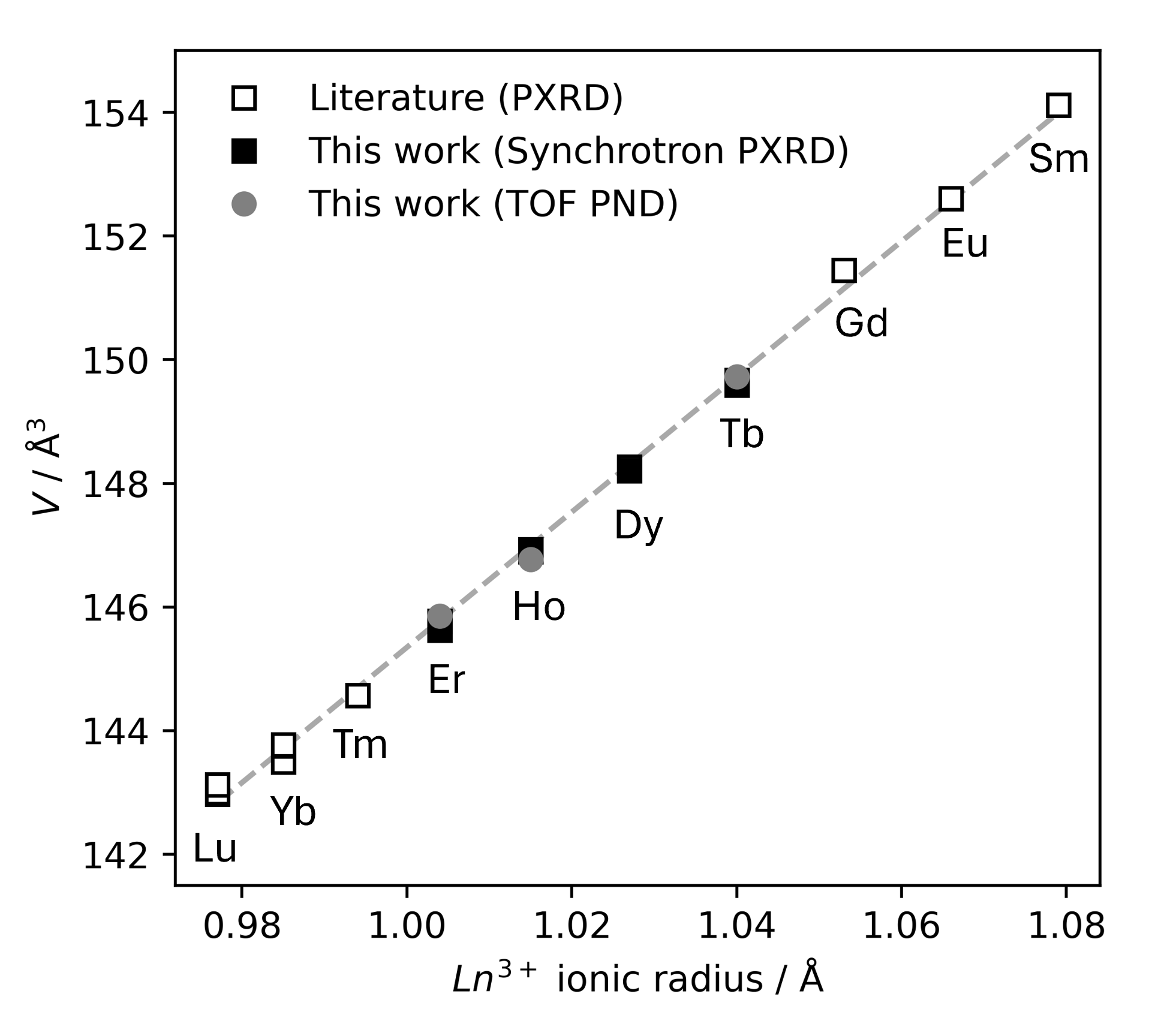}  
\caption{Unit cell volume of $M'$-\ch{\textit{Ln}TaO4} as a function of 8-coordinate $Ln^{3+}$ ionic radius \cite{Shannon1976} from Rietveld refinements against both high-resolution X-ray and neutron diffraction data. Error bars are smaller than the datapoints. Literature data taken from references \cite{Brixner1983,Mullens2023,Wang2017,Markiv2002,Kumar2024}.}
\label{fig:Mprime_volumes}
\end{figure}

Further structural characterisation was carried out using powder neutron diffraction (PND) on three samples, $M'$-\ch{TbTaO4}, \ch{HoTaO4} and \ch{ErTaO4} \cite{GEMdata}. Fig.~\ref{fig:rietveld_tb_pnd_rt} shows the Rietveld refinement for room-temperature data on $M'$-\ch{TbTaO4}; plots of the refinements for the other two compounds are presented in the Supplemental Material \cite{Supplemental}. Both PXRD and PND on the neutron samples showed small amounts ($\leq1$ wt \%) of impurities: \ch{\textit{Ln}3TaO7} \cite{Wakeshima2004,Mullens2024}; the competing $M$ phase of \ch{TbTaO4}; and \ch{Ta2O5} in the Ho sample, which unfortunately could not be eliminated after repeated firing of the larger neutron samples. However, all magnetometry and heat capacity measurements were done on smaller samples which were found to be phase-pure by high-resolution synchrotron PXRD. Unlike with PXRD, the relative neutron scattering lengths \cite{Sears1992} permitted refinement of the fractional coordinates of O ions as well as those of the metal ions in our samples, and the refined values (Table~\ref{table:lntao4_pnd}) matched those reported in the literature \cite{Hartenbach2005}. The unit cell volumes are  linearly related to the lanthanide ionic radius \cite{Shannon1976} and are consistent with literature reports, as shown by Fig.~\ref{fig:Mprime_volumes}.

\begin{table}[htbp]
\centering
\caption{Refined structural data for $M'$-\ch{\textit{Ln}TaO4} (\textit{Ln}~= Tb, Ho, Er) at room temperature from time-of-flight PND data (GEM, banks 3--6). Space group: $P2/c$. $Ln1$ atom at (0, $y_{Ln}$, 0.25); Ta1 atom at (0.5, $y_\mathrm{Ta}$, 0.75).}
\label{table:lntao4_pnd}
\resizebox{\columnwidth}{!}{
\begin{ruledtabular}
\begin{tabular}{c c c c c}
 & & Tb & Ho & Er \\
\midrule
\multicolumn{2}{c}{$a$ (\AA)} & 5.14258(2) & 5.11201(3) & 5.1007(15) \\
\multicolumn{2}{c}{$b$ (\AA)} & 5.49176(2) & 5.45320(3) & 5.4432(15) \\
\multicolumn{2}{c}{$c$ (\AA)} & 5.33763(2) & 5.29881(3) & 5.2864(15) \\
\multicolumn{2}{c}{$\beta$ (\degree)} & 96.6566(3) & 96.4658(5) & 96.4040(5) \\
\multicolumn{2}{c}{$V$ (\AA$^3$)} & 149.728(1) & 146.774(1) & 145.86(7) \\
\midrule
\textit{Ln}1 & $y$ & 0.73439(5) & 0.73414(7) & 0.73380(7) \\
Ta1 & $y$ & 0.80767(6) & 0.80476(7) & 0.80471(8) \\
O1 & $x$ & 0.74789(5) & 0.74879(7) & 0.74955(7) \\
   & $y$ & 0.41318(5) & 0.41620(7) & 0.41672(7) \\
   & $z$ & 0.39731(5) & 0.39987(8) & 0.40129(8) \\
O2 & $x$ & 0.27091(6) & 0.26904(8) & 0.26824(8) \\
   & $y$ & 0.06469(5) & 0.06442(7) & 0.06462(7) \\
   & $z$ & 0.49543(6) & 0.49575(8) & 0.49621(8) \\
$B_\mathrm{iso}$ (\AA$^2$) & $Ln$ & 0.229(4) & 0.148(5) & 0.328(7) \\
   & Ta & 0.218(4) & 0.200(6) & 0.281(7) \\
   & O & 0.430(3) & 0.418(4) & 0.498(5) \\   
\midrule
\multicolumn{2}{c}{$R_\mathrm{wp}$ (\%)} & 1.41 & 1.73 & 1.83 \\
\multicolumn{2}{c}{$\chi^2$} & 3.19 & 3.34 & 2.29 \\
\multicolumn{2}{c}{Impurities} & 0.47 wt \%\ \ch{Tb3TaO7}, & 2.7 wt \%\ \ch{Ho3TaO7}, & 0.8 wt \%\ \ch{Er3TaO7} \\
 & & 1.06 wt \%\ $M$-\ch{TbTaO4} & 1.1 wt \%\ \ch{Ta2O5} &  \\
\end{tabular}
\end{ruledtabular}
}
\end{table}

PND was also measured on the sample of $M'$-\ch{TbTaO4} at low temperatures using a cryostat. Upon cooling, the $a$, $b$ and $c$ lattice parameters decreased while the angle $\beta$ increased. Overall, the unit cell volume decreased by less than 0.5~\%\ between 300 and 1.5~K. The contraction of the unit cell is as expected for normal thermal expansion behaviour and no anomalous movements of the atomic positions were detected. The Supplemental Material \cite{Supplemental} includes a table of the refined lattice parameters, Table~S1, and plots of lattice parameters (Fig.~S9) and metal-oxygen bond lengths (Fig.~S10) as a function of $T$. At 1.5~K, large new peaks appeared which indicated long-range magnetic order in $M'$-\ch{TbTaO4}; the magnetic structure is discussed in Section~\ref{section:magpnd}.

\subsection{Magnetic measurements}

\begin{figure}[htbp]
\centering
\includegraphics[width=\textwidth]{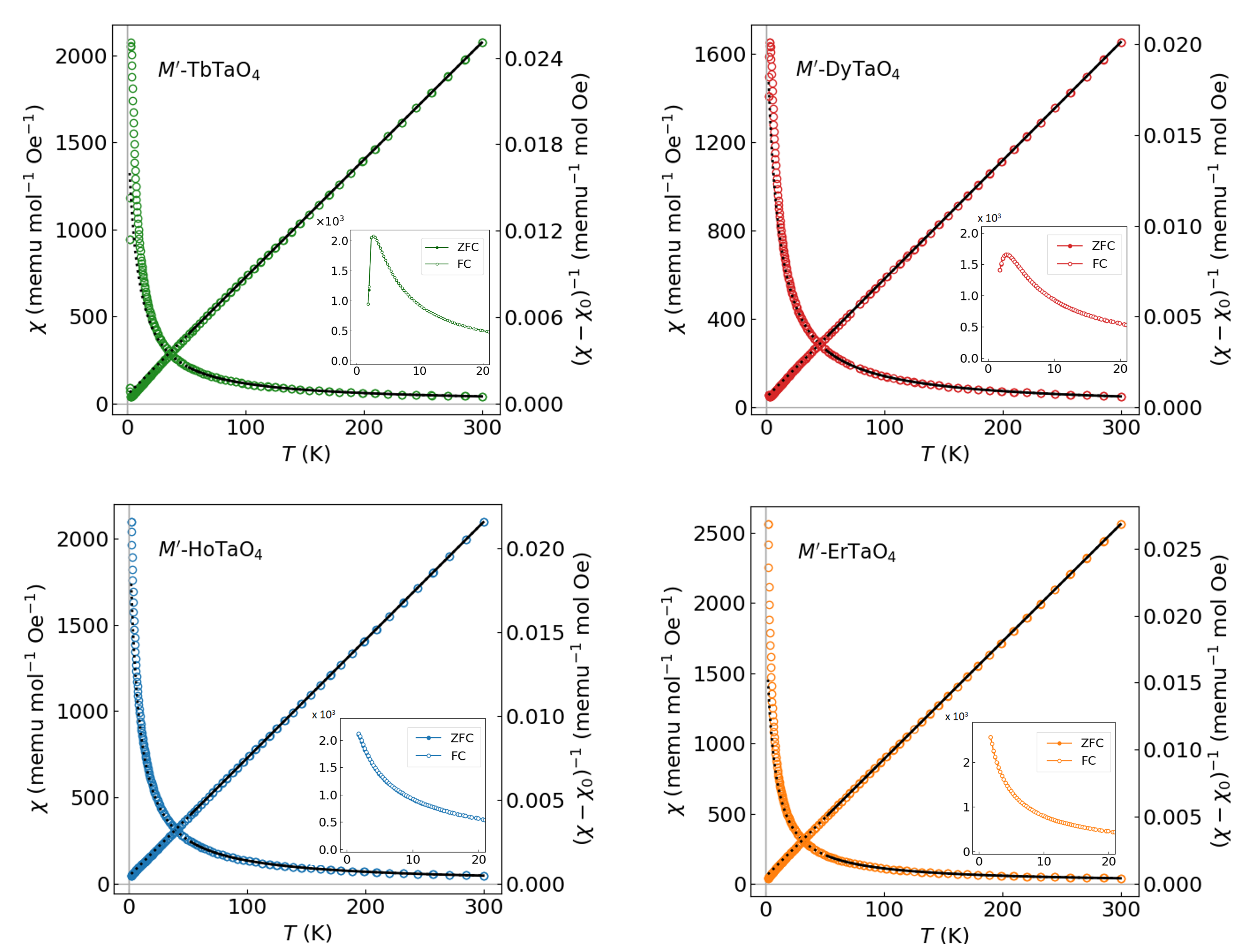}
\caption{Magnetic susceptibility $\chi$ and inverse susceptibility $(\chi-\chi_0)^{-1}$, fitted to the Curie-Weiss law in the range 50--300~K (black lines), for the four $M'$-\ch{\textit{Ln}TaO4} compounds in this study. The black dashed lines show the extrapolation of the fit below 50~K. Insets show the low-temperature region.}
\label{fig:curieweiss}
\end{figure}

Peaks in the magnetic susceptibility were visible for $M'$-\ch{TbTaO4} and $M'$-\ch{DyTaO4} at approximately 2.7~K for both samples (Fig.~\ref{fig:curieweiss}). There was no divergence between the zero-field-cooled (ZFC) and field-cooled (FC) susceptibility data which might have indicated spin-glass behaviour. This, combined with the shape of the peak for $M'$-\ch{TbTaO4} suggests a transition from paramagnetism to a long-range-ordered antiferromagnetic (AFM) state upon cooling. The peak for $M'$-\ch{DyTaO4} is significantly broader, which may indicate that below 2.7~K the magnetic moments have a short correlation length. Furthermore, the shape of the derivative $d(\chi T)/dT$ \cite{Fisher1962} shows a clear feature for $M'$-\ch{TbTaO4} but not for $M'$-\ch{DyTaO4} (Supplemental Material, Fig.~S12 \cite{Supplemental}), despite the peaks in $\chi(T)$ occurring at similar temperatures for the two compounds. Meanwhile, $M'$-\ch{HoTaO4} and $M'$-\ch{ErTaO4} show only Curie-Weiss behaviour with no susceptibility peaks visible in the temperature range measured.

The magnetic susceptibility $\chi=M/H$ was fitted to the modified Curie-Weiss law:
\begin{equation}
    \chi=\chi_0 + \frac{C}{T-\theta_{CW}}
    \label{eqn:cw}
\end{equation}
where $\chi_0$ is a small temperature-independent susceptibility, $C$ is the Curie constant ($C=\mu_{eff}^2/8$), and $\theta_\mathrm{CW}$ is the Curie-Weiss temperature. All chemical elements and compounds have a negative, temperature-independent diamagnetic contribution to the susceptibility, $\chi_\mathrm{Di}$, which can be estimated using Pascal's constants \cite{Bain2008} as $-8\times10^{-5}$ emu~mol$^{-1}$ for these four $M'$-\ch{\textit{Ln}TaO4} compounds. However, diamagnetism is a weak effect and is typically overridden by stronger magnetic effects such as paramagnetism \cite{Mugiraneza2022}, so lanthanide-containing solids are expected to have positive values of $\chi_0$. We used a Nelder-Mead optimisation method (Supplemental Material, Fig.~S13 \cite{Supplemental}) to find the best linear fit to the inverse susceptibility as a function of temperature. This approach minimises the scaled least-squares difference between $(T-\theta)/C$ and $(\chi-\chi_0)^{-1}$, which is more effective than regular Curie-Weiss fitting with free parameters on both sides of the equation (eqn \ref{eqn:cw}). This yielded small positive values of $\chi_0$ which are given in Table~\ref{table:curieweiss} along with the fitted values of $\theta_\mathrm{CW}$ and $\mu_\mathrm{eff}$. Fig.~\ref{fig:curieweiss} shows the Curie-Weiss fits as black solid lines with extrapolation below 50~K as dashed lines. The negative Curie-Weiss temperatures indicate antiferromagnetic interactions and the effective magnetic moments $\mu_{eff}$ agree quite well with the expected moments for free ions, $\mu=g_J\sqrt{J(J+1)}$.

\begin{table}[htbp]
\centering
\caption{Magnetic parameters for $M'$-\ch{\textit{Ln}TaO4} ($Ln=$ Tb, Dy, Ho, Er) from Curie-Weiss fitting in the range 50--300~K.}
\label{table:curieweiss}
\resizebox{\columnwidth}{!}{
\begin{ruledtabular}
\begin{tabular}{c c c c c}
 & $M'$-\ch{TbTaO4} & $M'$-\ch{DyTaO4} & $M'$-\ch{HoTaO4} & $M'$-\ch{ErTaO4} \\
\midrule
Sample mass (mg) & $6.2$ & $7.6$ & $6.8$ & $10.8$ \\
$\theta_\mathrm{CW}$ (K) & $-3.50(12)$ & $-6.75(7)$ & $-5.12(2)$ & $-3.93(5)$ \\
$\mu_\mathrm{eff}$ ($\mu_\mathrm{B}$) & $9.65(6)$ & $10.93(3)$ & $10.58(11)$ & $9.41(3)$ \\
Free-ion moment ($\mu_\mathrm{B}$) & $9.72$ & $10.65$ & $10.61$ & $9.58$ \\
$\chi_0$ (emu mol$^{-1}$ Oe$^{-1}$) & $2.0\times10^{-3}$ & $6.9\times10^{-4}$ & $1.3\times10^{-3}$ & $2.5\times10^{-3}$ \\
\end{tabular}
\end{ruledtabular}
}
\end{table}

\begin{figure}[htbp]
\centering
\includegraphics[width=0.5\textwidth]{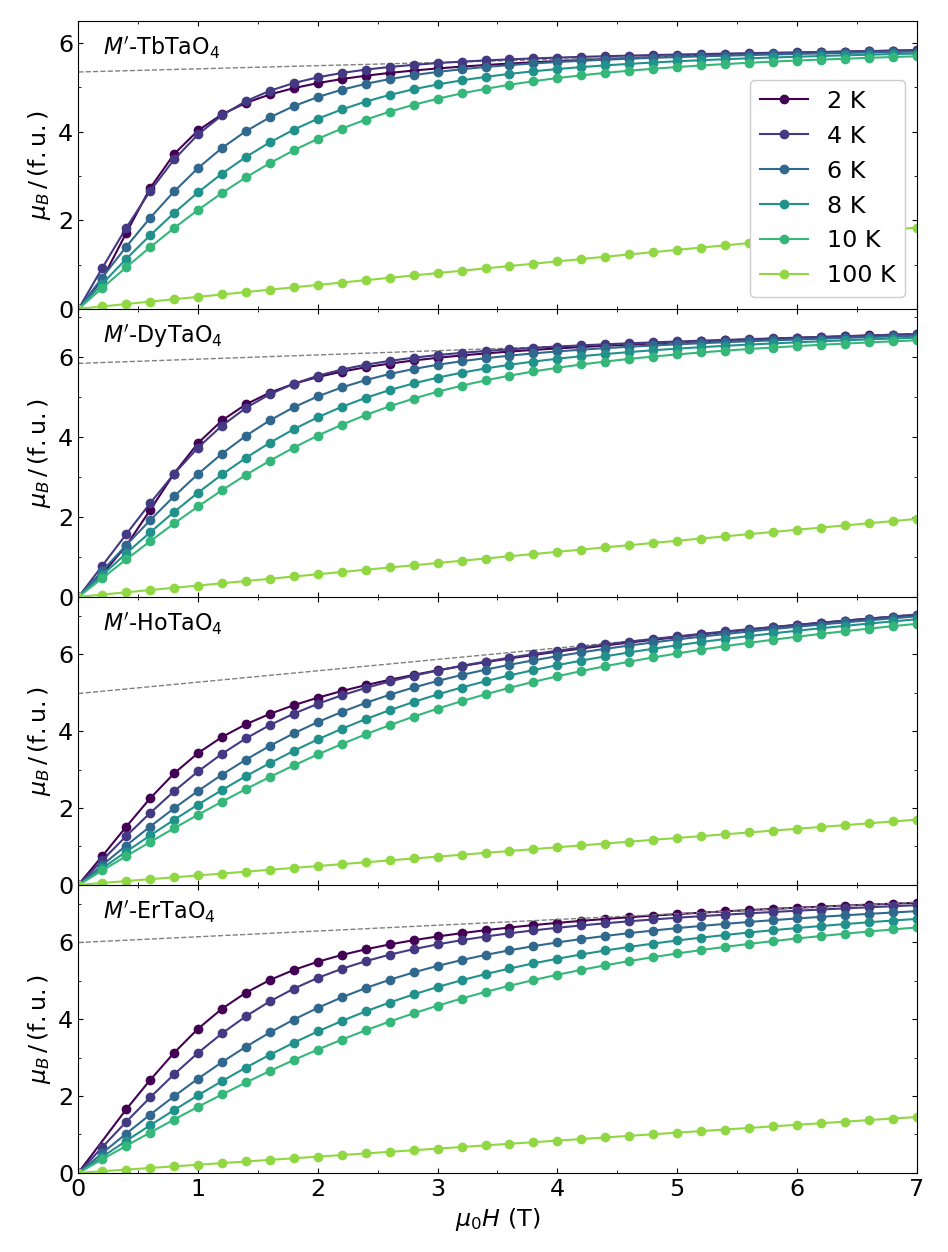}
\caption{Magnetic isotherms for $M'$-\ch{\textit{Ln}TaO4}. The dashed lines show the linear fits in the region 5-7~T, extrapolated back to 0~T.}
\label{fig:MHs}
\end{figure}

Magnetic isotherms were measured for all samples at $T=$ 2, 4, 6, 8, 10 and 100~K and are shown in Fig.~\ref{fig:MHs}. At $T\leq 10$~K and at high fields, the magnetisation saturates to a straight line representing the underlying paramagnetism or crystal electric field (CEF) effects. The data at $T=2$~K were therefore fitted to a linear function between 5 and 7~T and the line was extrapolated back to zero field (grey dashed line) to obtain an estimate for the saturation magnetisation, $M_\mathrm{sat}$. The magnitude of $M_\mathrm{sat}$ can be compared with the predicted value ($g_J.J$ for Heisenberg, $g_J.J/2$ for Ising, or $2g_J.J/3$ for XY anisotropy, in powder samples \cite{Bramwell2000}) to give an indication of the single-ion anisotropy in these systems. The data indicate that all compounds exhibit significant single-ion anisotropy, and in particular that $M'$-\ch{HoTaO4} may be an Ising system ($M_\mathrm{sat}=5$ and $g_J.J=10$~$\mu_\mathrm{B}$/f.u.) and $M'$-\ch{ErTaO4} an XY system ($M_\mathrm{sat}=6$ and $g_J.J=9$~$\mu_\mathrm{B}$/f.u.). However, errors arising from the extrapolation and/or the sample mass mean that caution should be exercised in drawing these conclusions from the $M(H)$ data alone. For example, the magnitude of $M_\mathrm{sat}$ for $M'$-\ch{TbTaO4} lies somewhat between the expected values for Ising and XY anisotropy, yet the neutron diffraction data (Section~\ref{section:magpnd}) indicate an almost easy-axis (Ising) behaviour. Additional neutron diffraction down to lower temperatures on the Ho and Er samples, and/or an isotopically enriched Dy sample, would be required to confirm the type of anisotropy. Spectroscopic measurements to determine the CEF levels would also provide information on anisotropy.

\subsection{\label{section:hc}Specific heat}
\begin{figure}[htbp]
\centering
\includegraphics[width=0.5\textwidth]{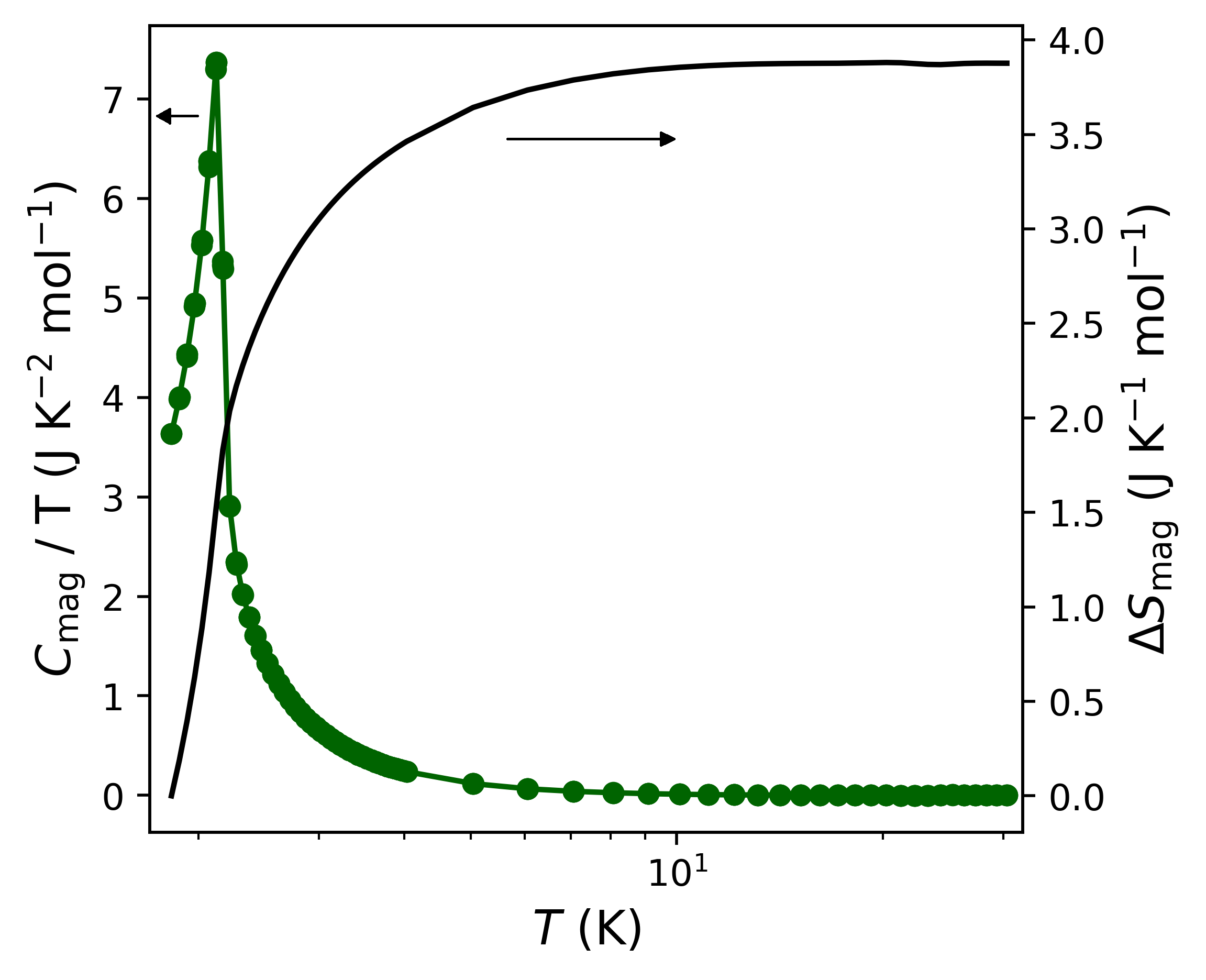}
\caption{$M'$-\ch{TbTaO4} magnetic specific heat data (left axis) and magnetic entropy change (right axis).}
\label{fig:Tb_HC}
\end{figure}

\begin{figure}[htbp]
\centering
\includegraphics[width=\textwidth]{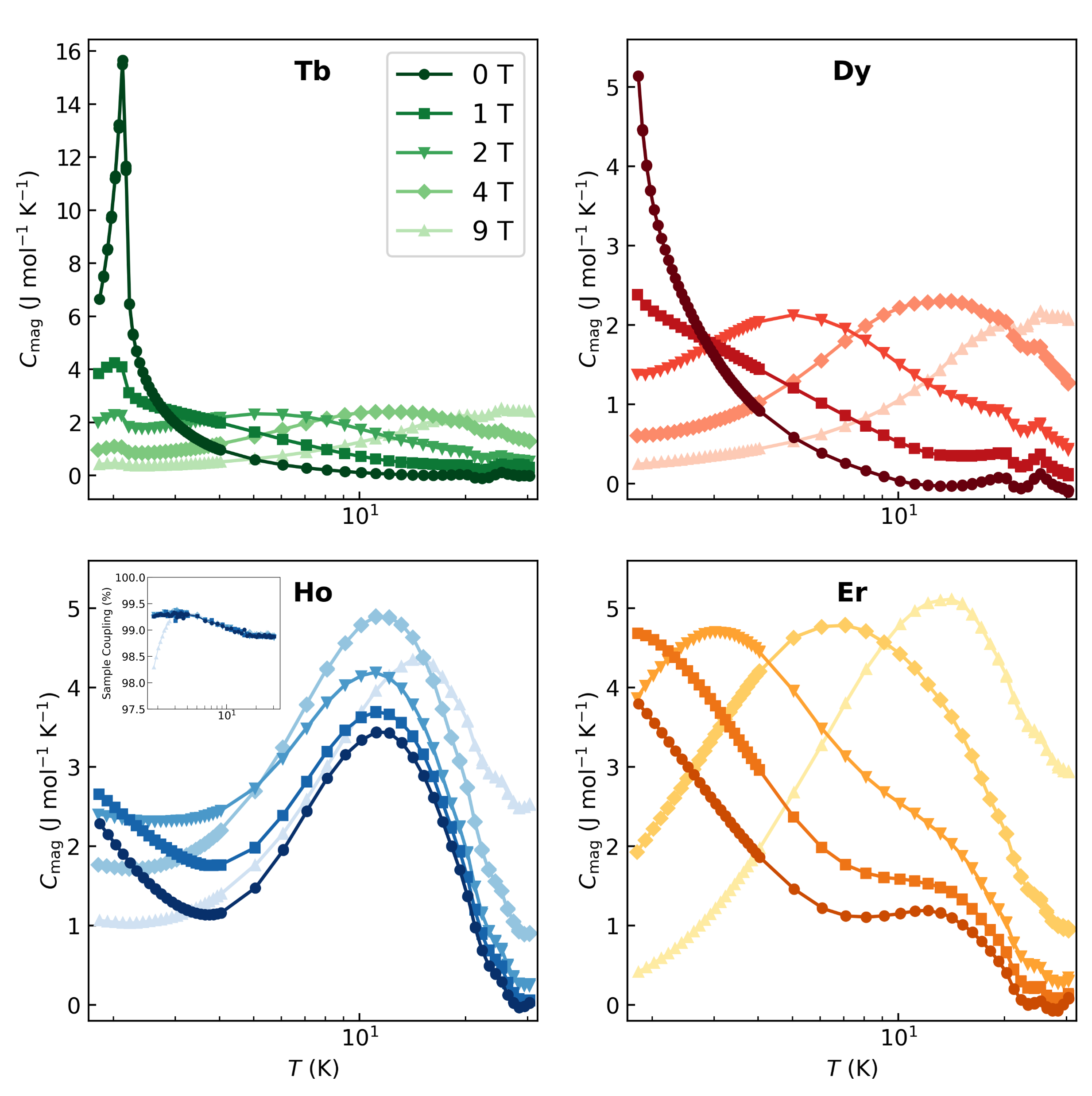}
\caption{Magnetic specific heat data at various applied fields for the four $M'$-\ch{\textit{Ln}TaO4} samples. Inset for $Ln=$ Ho shows the sample coupling.}
\label{fig:HC_all}
\end{figure}

The zero-field magnetic specific heat for $M'$-\ch{TbTaO4} is given in Fig.~\ref{fig:Tb_HC}. A sharp $\lambda$-type transition is observed at $T=2.1$~K, indicating long-range AFM ordering. The temperature of the peak in heat capacity is lower than that observed in the magnetic susceptibility (2.7~K), which is similar behaviour to the other polymorph of this compound, $M$-\ch{TbTaO4} \cite{Kelly2022a}. In addition, a peak in $d(\chi T)/dT$ occurs at the same temperature as the peak in the specific heat (Supplemental Material, Fig.~S12 \cite{Supplemental}), which is typical for antiferromagnets \cite{Fisher1962}. Integration of $C_\mathrm{mag}/T$ yields the magnetic entropy, $\Delta S_\mathrm{mag}$, which approaches 4~J~K$^{-1}$~${\mathrm{mol}_\mathrm{Tb}}^{-1}$ at 30~K. This value is less than 20~\%\ of the expected $\Delta S_\mathrm{mag}=R\ln{(2J+1)}=21.32$~J~K$^{-1}$~mol$^{-1}$ ($J=6$); clearly, since the lower-temperature side of the $\lambda$-like peak is incomplete, there is still significant further entropy to be lost on cooling the material below 1.8~K. The specific heat in zero field and applied fields (1, 2, 4 and 9~T) is also shown in Fig.~\ref{fig:HC_all}, top right. Upon application of a magnetic field, the sharp peak is suppressed and becomes a broad feature which moves to higher temperatures with increasing field.

In contrast, for $M'$-\ch{DyTaO4} no such sharp peak is observed (Fig.~\ref{fig:HC_all}, top right), which agrees with the absence of a peak in $d(\chi T)/dT$ \cite{Fisher1962} (Supplemental Material, Fig.~S12 \cite{Supplemental}). There is no evidence for either long-range order or spin freezing above 1.8~K in either the zero-field or 1~T specific heat data. At 2~T we observe a broad peak, centred around 5~K. This peak then shifts to higher temperatures as the field increases, which is typical of a Schottky anomaly. Such behaviour arises from Zeeman splitting of a Kramers doublet state with $J_\mathrm{eff}=\frac{1}{2}$, as observed in the isostructural $M'$-\ch{YbTaO4} at all fields \cite{Kumar2024}. However, the absence of a peak in the lower-field datasets suggests that the ground state at zero field is not a Kramers doublet, and that there is a field-induced transition between 1 and 2~T, perhaps reflecting suppression of the short-range order suggested by the magnetic susceptibility. In Fig.~\ref{fig:MHs} the magnetisation curves for $M'$-\ch{DyTaO4} at 2~K and 4~K cross over twice at low fields, which also hints at magnetic ordering. There are also two small anomalies in the $M'$-\ch{DyTaO4} data around $T=20$ and 25~K whose origin is unknown. They may be an artefact of measurements or related to the sample environment, because less pronounced features appear at the same temperatures in the \ch{HoTaO4} and \ch{ErTaO4} data and they are independent of the applied field.

Like Dy$^{3+}$ ions, Er$^{3+}$ ions have an odd number of electrons and are therefore also Kramers ions. $M'$-\ch{ErTaO4} displays a Schottky-like peak similar to that of $M'$-\ch{DyTaO4} (Fig.~\ref{fig:HC_all}, bottom right) but, unlike \ch{DyTaO4}, this behaviour appears to extend to the lowest fields. There is a second broad feature between around 7 and 22~K in the zero-field data, which seems to be suppressed in applied magnetic fields and may represent short-range magnetic correlations; the feature might be investigated in the future using neutron diffraction and/or inelastic scattering in this temperature range.

Finally, the specific heat data for $M'$-\ch{HoTaO4} are shown in Fig.~\ref{fig:HC_all} (bottom left). A broad hump centred around 8~K is clearly visible and, with the exception of the 9~T data, the temperature of the maximum remains fairly independent of applied field, unlike in the Dy and Er analogues. Broad features like this have been observed in other compounds containing Ho$^{3+}$ (a non-Kramers ion), especially when the Ho$^{3+}$ ions are in low-symmetry crystallographic environments; in $\pi$-\ch{HoBO3} the behaviour was ascribed to a non-magnetic singlet ground state, with the broad feature arising from van Vleck paramagnetism \cite{Mukherjee2018a}. In $M'$-\ch{HoTaO4}, the Ho$^{3+}$ ion is on the $2e$ Wyckoff position of space group 13 ($P2/c$) with site symmetry $C_2$. The low site symmetry and non-Kramers nature of Ho$^{3+}$ mean that its $2J+1=15$-fold degeneracy may be lifted by interaction with the surrounding CEF, i.e.~there is no symmetry-enforced ground state doublet, unlike in many Er$^{3+}$ or Yb$^{3+}$ compounds. Hence, we suggest that the specific heat data on $M'$-\ch{HoTaO4} represent a singlet ground state. The higher-temperature peak in the 9~T dataset is probably an anomaly, related to some slippage of the sample platform or thermocouples between the 4~T and 9~T measurements, since there is no evidence in the $M(H)$ data for a field-induced transition in this range. The inset to Fig.~\ref{fig:HC_all} shows how the sample coupling for $M'$-\ch{HoTaO4} is different for the 9~T data at low temperatures, supporting this hypothesis; for the other three compounds, the sample coupling had a slight $T$-dependence but was independent of applied field.

\subsection{\label{section:magpnd}Magnetic structure of $M'$-\ch{TbTaO4}}
The magnetic structure of $M'$-\ch{TbTaO4} was investigated using PND \cite{GEMdata} below the ordering temperature of $T_\mathrm{N}=2.1$~K. Additional peaks appeared between the 30~K and 1.5~K datasets in positions which indicated commensurate magnetic ordering. Using the FullProf Suite \cite{Rodriguez-Carvajal1993}, a suitable magnetic ordering vector was found to be $\vec{k}=(\frac{1}{2},\frac{1}{2},0)$. The magnetic peaks were then modelled in TOPAS v5 \cite{Coelho2018} in magnetic space group 14.80 ($P_a2_1/c$). Each Tb$^{3+}$ ion carries a magnetic moment with refined components $-0.702(8)$, 0 and $8.053(6)$~$\mu_\mathrm{B}$ along the $a$, $b$ and $c$ axes respectively. This corresponds to $-1.652(12)$, 0 and $7.996(6)$~$\mu_\mathrm{B}$ in Cartesian axes and an ordered moment magnitude of $8.17(9)$~$\mu_\mathrm{B}$. This is consistent with the expectation of $g_JJ=9$~$\mu_\mathrm{B}$ for Tb$^{3+}$ ions. Note that this magnetic space group does permit a non-zero $y$-component to the magnetic moment. However, its value always refined to zero (from different trial starting points) with a large uncertainty reported by the TOPAS software, so it was subsequently fixed at zero. Thus, in this model, the magnetic moments are primarily directed along the $c$-direction and are fractionally canted towards $a$. The nearest-neighbour magnetic moments are coupled antiferromagnetically within the $bc$ plane, and the next-nearest-neighbour interactions are ferromagnetic, creating a zigzag-like ordering pattern. Adjacent layers along the $a$ axis are coupled antiferromagnetically.

\begin{figure}[htbp]
\centering
\includegraphics[width=\textwidth]{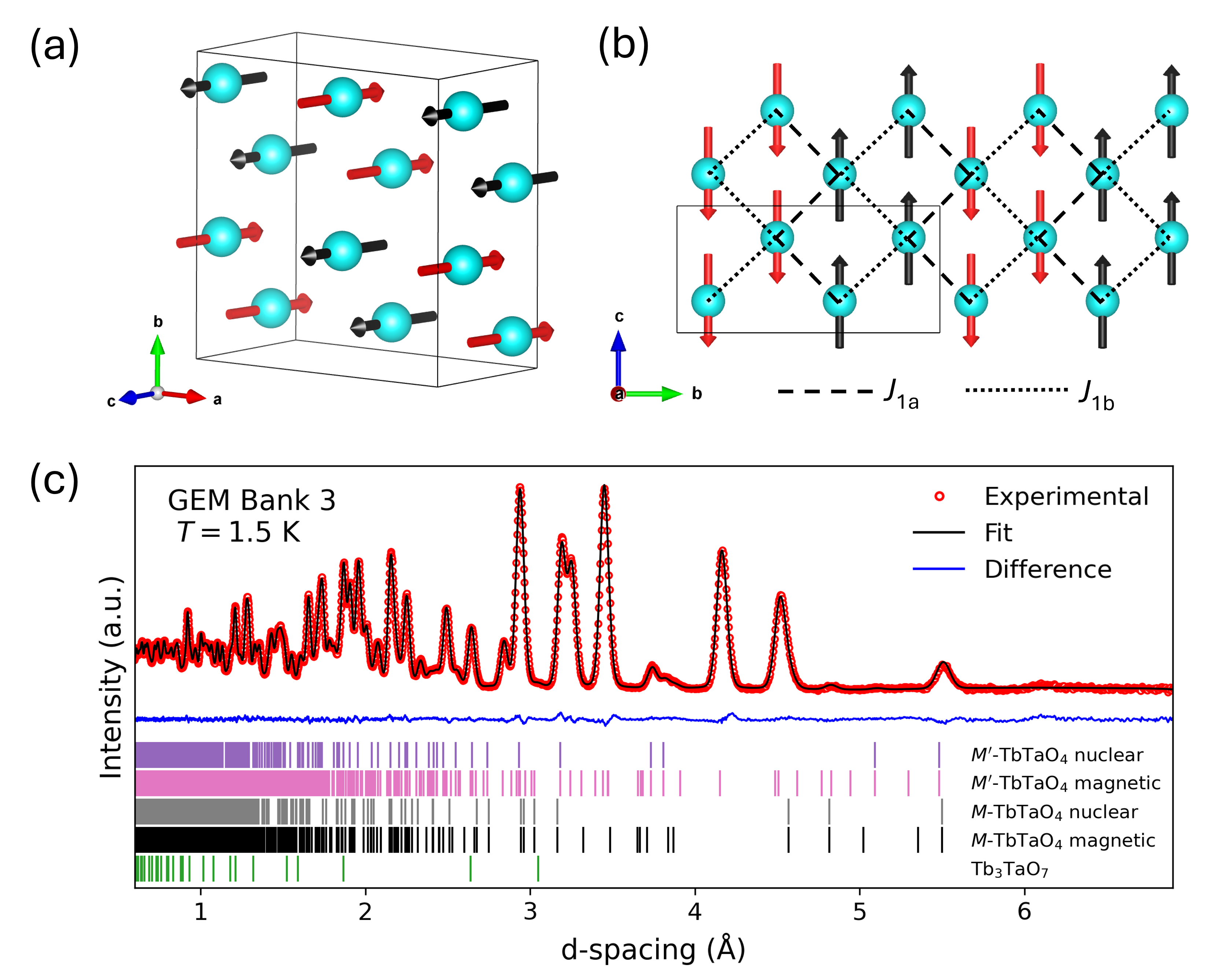}
\caption{(a) Magnetic structure model for one magnetic unit cell of $M'$-\ch{TbTaO4} at 1.5~K. (b) Magnetic structure model for a single layer (distorted square lattice). Dashed lines are the nearest-neighbour distances at 3.7054(4)~\AA, dotted lines are the next-nearest-neighbour distances at 3.9493(4)~\AA. (c) Rietveld refinement against TOF PND data at $T=1.5$~K: GEM bank 3.}
\label{fig:rietveld_tb_1p5K}
\end{figure}

The magnetic structure model for $M'$-\ch{TbTaO4} is shown in Fig.~\ref{fig:rietveld_tb_1p5K}(a) and (b), and the Rietveld refinement at 1.5~K (bank 3) in Fig.~\ref{fig:rietveld_tb_1p5K}(c). Rietveld refinements for all banks are available in the Supplemental Material, Figure S11 \cite{Supplemental}, along with a table giving the structural parameters for a one-phase combined nuclear and magnetic structure, Table S2. In addition to the $M'$-\ch{TbTaO4}, two other minor phases contribute intensity to the 1.5~K neutron diffraction pattern: 1.06 wt~\%\ $M$-\ch{TbTaO4} (both nuclear and magnetic reflections \cite{Kelly2022a}), and 0.47 wt~\%\ \ch{Tb3TaO7}. \ch{Tb3TaO7} is known to undergo two magnetic ordering transitions at 2.9~K and 3.6~K \cite{Wakeshima2004}, which may be the origin of the very small regions of unindexed intensity in the 1.5~K data. However, the magnetic structure of \ch{Tb3TaO7} has not been reported and could not be deduced from our data because of the very low weight fraction of \ch{Tb3TaO7} in this sample.

In comparison to the $M'$ phase described above, the $M$ polymorph of \ch{TbTaO4}, which has a three-dimensional ``stretched diamond'' network of Tb$^{3+}$ ions, displays AFM order below $T_\mathrm{N}=2.25$~K with the moments directed along the $a$-axis \cite{Kelly2022a}. Both phases of \ch{TbTaO4} display almost easy-axis anisotropy and antiferromagnetic coupling between nearest-neighbour Tb ions. However, differences in their crystal structures lead to important differences in the magnetic structures. In the $M$ phase, there are four near-neighbour AFM interactions at similar distances ($2\times J_{1a}$ at 3.75~\AA, $2\times J_{1b}$ at 3.84~\AA) and the closest ferromagnetically coupled pair of spins are much further apart at $\sim5.05$~\AA. In contrast, the $M'$ phase has AFM $2\times J_{1a}$ at 3.71~\AA\ and FM $2\times J_{1b}$ at 3.95~\AA, plus AFM inter-layer coupling $J_2$ (5.13~\AA). The tilting of the magnetic moments out of the plane in the $M'$ phase might be a way to partially relieve the frustration that arises from these competing AFM and FM near-neighbour interactions. The experimental $T_\mathrm{N}$ is slightly lower for the $M'$ phase than the $M$ phase (2.1 vs 2.25 K). However, the Curie-Weiss temperature is smaller for $M'$ than $M$, and the estimated frustration parameter $f=|\theta_\mathrm{CW}/T_\mathrm{N}|$ is therefore also smaller, although both values are modest ($M$: $f\approx4.2$, $M'$: $f\approx1.7$). We can use mean-field estimates of the nearest-neighbour exchange ($J_\mathrm{nn}$) and dipolar ($D_\mathrm{nn}$) interactions, equations \ref{equation:exchange} and \ref{equation:dipolar}, to rationalise the difference in magnetic ordering temperatures:

\begin{equation}
    J_\mathrm{nn}=\frac{3k_\mathrm{B}|\theta_\mathrm{CW}|}{2nJ(J+1)}
    \label{equation:exchange}
\end{equation}

\begin{equation}
    D_\mathrm{nn}=\frac{\mu_0\mu_\mathrm{eff}^2}{4\pi R^3 J(J+1)}
    \label{equation:dipolar}
\end{equation}

where $n$ is the number of ``nearest neighbours'' and $J$ is the total angular momentum quantum number, $J=6$ for Tb$^{3+}$. We took $n=4$ for both structures; for $R_\mathrm{nn}$ we used an average of the four closest Tb--Tb distances, which are within 7~\%\ of each other (compared with the next closest ion, which is at a distance more than 30~\%\ larger than $R_\mathrm{nn}$ in both phases). In $M'$-\ch{TbTaO4}, its smaller Curie-Weiss temperature and greater average Tb--Tb separation lead to smaller values for both the superexchange and dipolar interactions. These weaker interactions are reflected in the slightly lower magnetic ordering temperature. Table~\ref{table:interactions} gives the relevant parameters and estimated interaction strengths for all four samples of $M'$-\ch{\textit{Ln}TaO4} with comparisons to the $M$ polymorphs \cite{Kelly2022a}. For each pair of compounds with the same \textit{Ln}, the values of $J_\mathrm{nn}$ and $D_\mathrm{nn}$ are smaller for the $M'$ phase than for the $M$ phase. We note that the difference in $D_\mathrm{nn}$ for $M'$ versus $M$ gets smaller across the lanthanide series (as the average \textit{Ln}--\textit{Ln} distance, $R$, gets smaller), until $D_\mathrm{nn}$ is the same for $M$- and $M'$-\ch{ErTaO4}. This demonstrates an interesting correlation with the minimum \textit{Ln}$^{3+}$ radius, below which the $M$ phase can no longer be synthesised under ambient pressure conditions \cite{Brixner1983,M-YbTaO4}. The use of pressure to produce compounds (magnetic or non-magnetic) which are inaccessible by conventional synthesis methods is well documented \cite{Solana-Madruga2021} but typically limited to small volumes of polycrystalline samples. However, the application of a magnetic field during synthesis has lately been revealed as another possible control parameter for synthesis. Under the so-called magneto-synthesis, large single crystals of \ch{BaIrO3} were produced with different structures and magnetic behaviour depending on the field applied during crystal growth \cite{Cao2026}. The lanthanide tantalates, with their range of different crystal structures and magnetic properties, may be good candidates for future investigations into the wider applicability of this technique.

\begin{table}[htbp]
\centering
\caption{Estimates of the nearest-neighbour exchange and dipolar interactions for $M'$- and $M$-\ch{\textit{Ln}TaO4} ($Ln=$ Tb, Dy, Ho, Er). The Curie-Weiss temperatures and effective magnetic moments for the $M$ phase are taken from ref.~\citenum{Kelly2022a}.}
\label{table:interactions}
\resizebox{\columnwidth}{!}{
\begin{ruledtabular}
\begin{tabular}{c c c c c c c c c}
 & \multicolumn{2}{c}{\ch{TbTaO4}} & \multicolumn{2}{c}{\ch{DyTaO4}} & \multicolumn{2}{c}{\ch{HoTaO4}} & \multicolumn{2}{c}{\ch{ErTaO4}} \\
 & $M$ & $M'$ & $M$ & $M'$ & $M$ & $M'$ & $M$ & $M'$ \\
\midrule
$\theta_\mathrm{CW}$ (K) & $-9.49$ & $-3.50$ & $-6.88$ & $-3.0$ & $-7.84$ & $-6.38$ & $-7.43$ & $-6.05$ \\
$\mu_\mathrm{eff}$ ($\mu_\mathrm{B}$) & $10.18$ & $9.65$ & $10.71$ & $10.64$ & $10.57$ & $10.66$ & $9.44$ & $9.55$ \\
$n$ & 4 & 4 & 4 & 4 & 4 & 4 & 4 & 4 \\
$J$ & 6 & 6 & 7.5 & 7.5 & 8 & 8 & 7.5 & 7.5 \\
$R_\mathrm{nn}$ (\AA) & 3.799 & 3.830 & 3.786 & 3.816 & 3.775 & 3.805 & 3.763 & 3.794 \\
$J_\mathrm{nn}$ (K) & 0.0847 & 0.0313 & 0.0405 & 0.0176 & 0.0408 & 0.0332 & 0.0437 & 0.0356 \\
$D_\mathrm{nn}$ (K) & 0.0280 & 0.0246 & 0.0207 & 0.0199 & 0.0180 & 0.0179 & 0.0163 & 0.0163 \\
\end{tabular}
\end{ruledtabular}
}
\end{table}

Notably, both phases of \ch{TbTaO4} are the only compounds in their respective $M'$-\ch{\textit{Ln}TaO4} ($Ln=$ Tb--Yb) or $M$-\ch{\textit{Ln}TaO4} series ($Ln=$ Nd--Yb) which have been found to display long-range order above 1.8~K. Ramanathan \textit{et al.} reported that $M'$-\ch{YbTaO4} has no long-range order down to 0.1~K, but to our knowledge no other compounds in this family have been studied below 1.8~K. Further low-temperature experiments on the other \ch{\textit{Ln}TaO4} compounds should be carried out where possible, to compare and contrast between different lanthanide ions and to determine whether there is any correlation between the N\'{e}el temperatures or magnetic structures of the $M$ and $M'$ phases for lanthanide ions other than Tb$^{3+}$.

\section{\label{section:conclusion}Conclusions}
We have investigated the crystal structure and bulk magnetic properties of the lanthanide tantalum oxides, $M'$-\ch{\textit{Ln}TaO4}, where $Ln=$ Tb, Dy, Ho and Er. Powder X-ray and neutron diffraction were used to study the crystal structure. Using magnetometry and heat capacity measurements, we observe LRO at $T_\mathrm{N}=2.1$~K for \ch{TbTaO4} and no LRO above 1.8~K for the compounds where $Ln=$ Dy, Ho or Er, although the susceptibility of $M'$-\ch{DyTaO4} suggests possible short-range magnetic ordering below $T\approx2.7$~K. $M'$-\ch{ErTaO4} has a Kramers doublet ground state as evidenced by the Schottky-type anomaly in field-dependent specific heat data, similar to the previously reported $M'$-\ch{YbTaO4}. A similar feature was seen for $M'$-\ch{DyTaO4} above 2~T, but there was no peak in the heat capacity at lower fields. This behaviour could be related to short-range order and/or CEF effects.

Powder neutron diffraction indicates that the magnetic moments in $M'$-\ch{TbTaO4} below $T_\mathrm{N}$ are directed primarily along the $c$-axis in a long-range antiferromagnetic structure with $\vec{k}=(\frac{1}{2},\frac{1}{2},0)$, with an ordered magnitude of 8.17(9)~$\mu_\mathrm{B}$ at 1.5~K. We compared the magnetic ordering temperatures of the $M$ and $M'$ polymorphs of \ch{TbTaO4} and related these to the estimated strengths of superexchange and dipolar coupling between Tb$^{3+}$ ions. For all $Ln$, the interactions are weaker for the $M'$ phase than for the $M$ phase, which correlates with the smaller $T_\mathrm{N}$ for $M'$- vs $M$-\ch{TbTaO4}. This work provides insight into how the magnetic anisotropy and ordering behaviour of rare-earth oxides may be significantly affected, either by substitution of different magnetic ions into a given crystal structure, or by differences in the crystal structure for a given magnetic ion (specifically Tb$^{3+}$). Future investigations at sub-kelvin temperatures will enable us to extend this work to other lanthanide ions.

\begin{acknowledgments}
We acknowledge funding from the EPSRC for the use of the Advanced Materials Characterisation Suite (EP/M000524/1). N.D.K.\ acknowledges funding from Jesus College, Cambridge for a Research Fellowship. S.E.D.\ acknowledges funding from the EPSRC (EP/T028580/1). We acknowledge I11 beamline at the Diamond Light Source, UK, for the synchrotron XRD measurement done under BAG proposal CY40166-2, and thank Farheen N.\ Sayed for organising the beamtime. Data collection at the ISIS Neutron and Muon Source was supported by a beamtime allocation RB2520021 from the Science and Technology Facilities Council \cite{GEMdata}. We thank James M.~A.\ Steele and Jeongjae Lee for assistance with neutron data collection.

Research data supporting the findings of this study will be made openly available at https://doi.org/10.17863/CAM.125160 \cite{Data}.

\end{acknowledgments}

\bibliography{library}

\appendix


\renewcommand{\thefigure}{S\arabic{figure}}
\setcounter{figure}{0}
\renewcommand{\thetable}{S\arabic{table}}
\setcounter{table}{0}

\newpage
\section{\label{app:rietveld}PXRD Rietveld refinements}

\begin{figure}[htbp]
\centering
\includegraphics[width=0.8\textwidth]{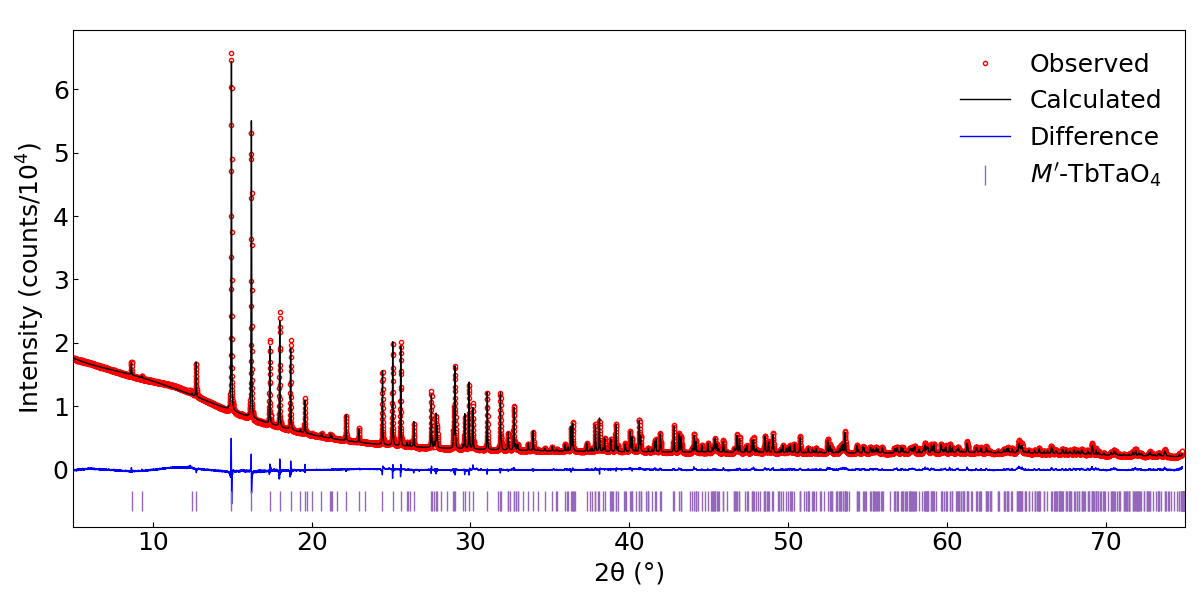}
\caption{Rietveld refinement against room-temperature synchrotron PXRD data ($\lambda= 0.82869$~\AA) for $M'$-\ch{TbTaO4}. Red circles: observed data, black line: calculated pattern, blue line: difference pattern, purple tick marks: Bragg reflection positions.}
\label{fig:rietveld_tb}
\end{figure}

\begin{figure}[htbp]
\centering
\includegraphics[width=0.8\textwidth]{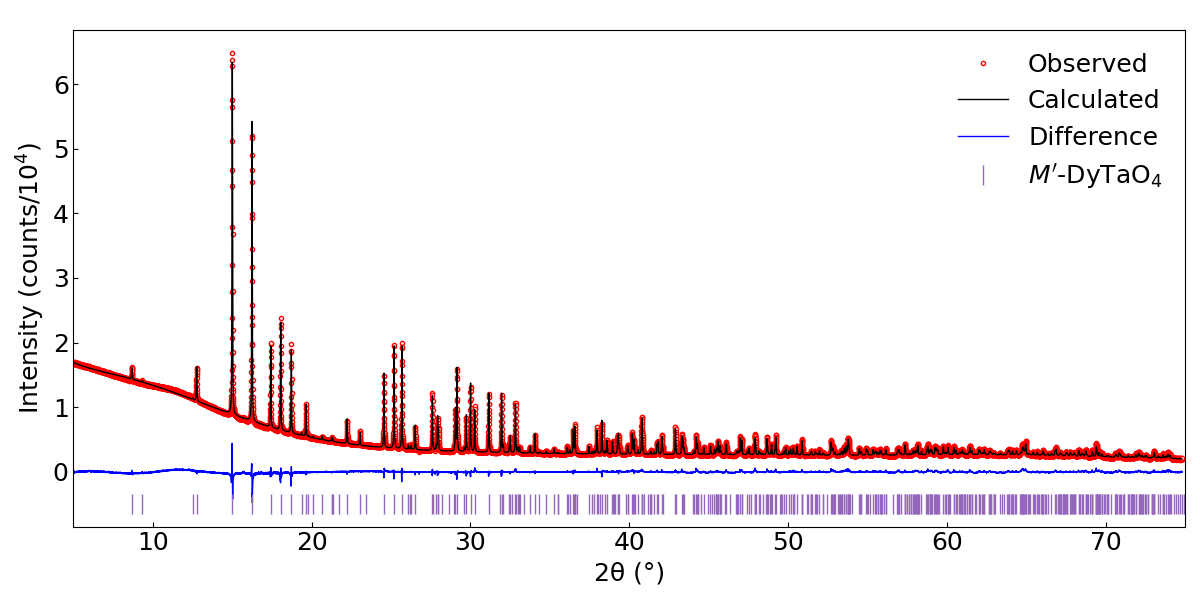}
\caption{Rietveld refinement against room-temperature synchrotron PXRD data ($\lambda= 0.82869$~\AA) for $M'$-\ch{DyTaO4}. Red circles: observed data, black line: calculated pattern, blue line: difference pattern, purple tick marks: Bragg reflection positions.}
\label{fig:rietveld_dy}
\end{figure}

\begin{figure}[htbp]
\centering
\includegraphics[width=0.8\textwidth]{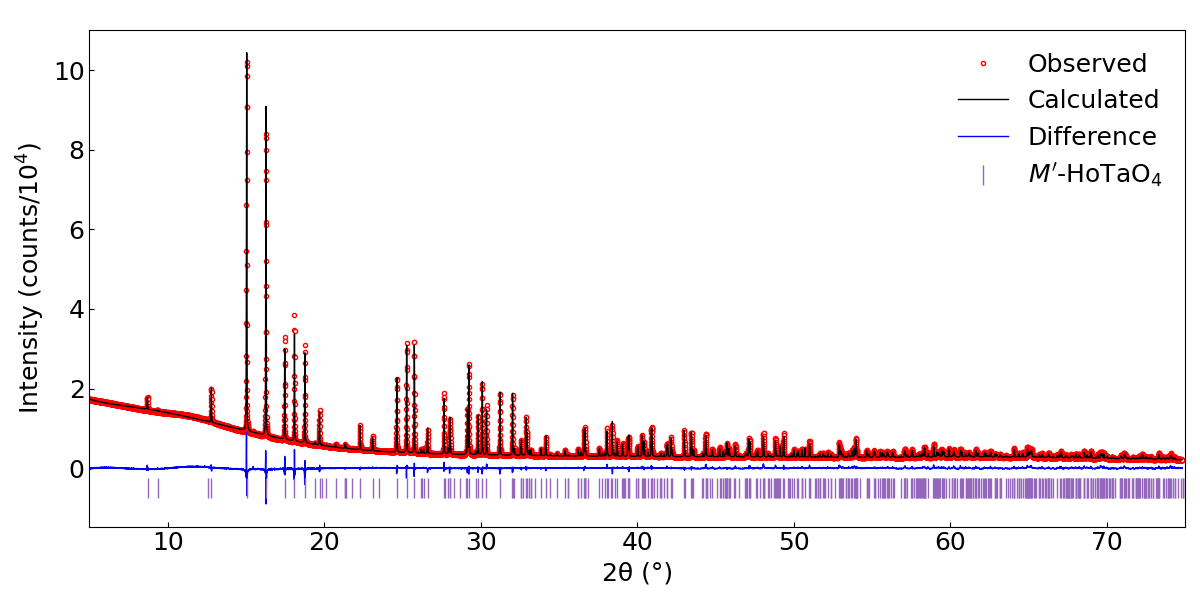}
\caption{Rietveld refinement against room-temperature synchrotron PXRD data ($\lambda= 0.82869$~\AA) for $M'$-\ch{HoTaO4}. Red circles: observed data, black line: calculated pattern, blue line: difference pattern, purple tick marks: Bragg reflection positions.}
\label{fig:rietveld_ho}
\end{figure}

\begin{figure}[htbp]
\centering
\includegraphics[width=0.8\textwidth]{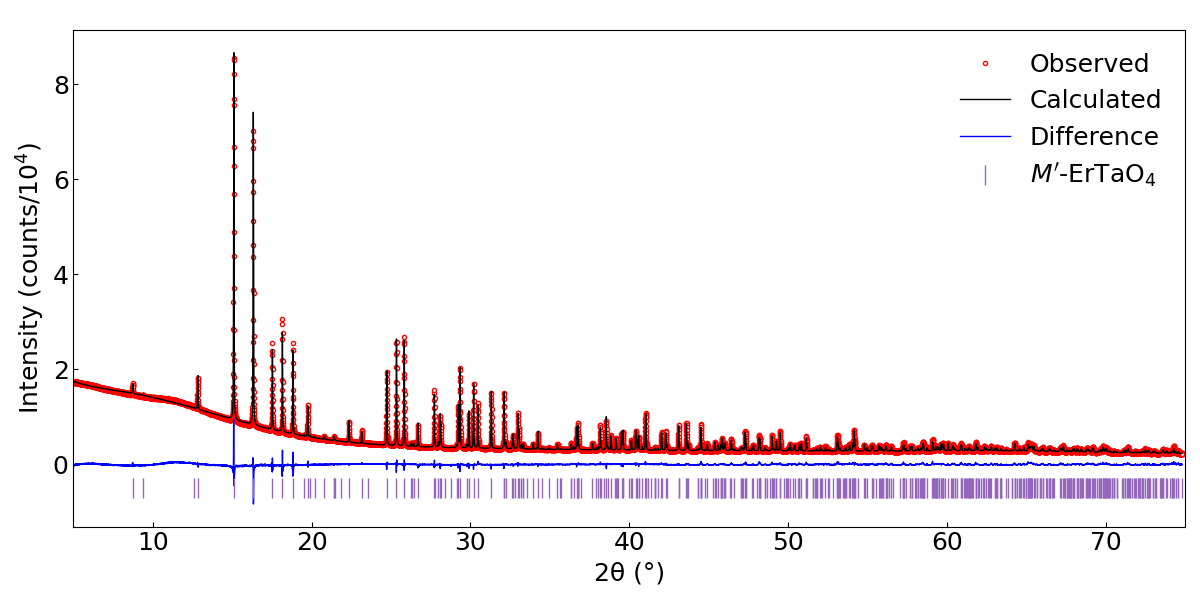}
\caption{Rietveld refinement against room-temperature synchrotron PXRD data ($\lambda= 0.82869$~\AA) for $M'$-\ch{ErTaO4}. Red circles: observed data, black line: calculated pattern, blue line: difference pattern, purple tick marks: Bragg reflection positions.}
\label{fig:rietveld_er}
\end{figure}

\clearpage

\section{\label{app:pnd}Neutron refinements and structural parameters}

\begin{figure}[htbp]
\centering
\includegraphics[width=0.8\textwidth]{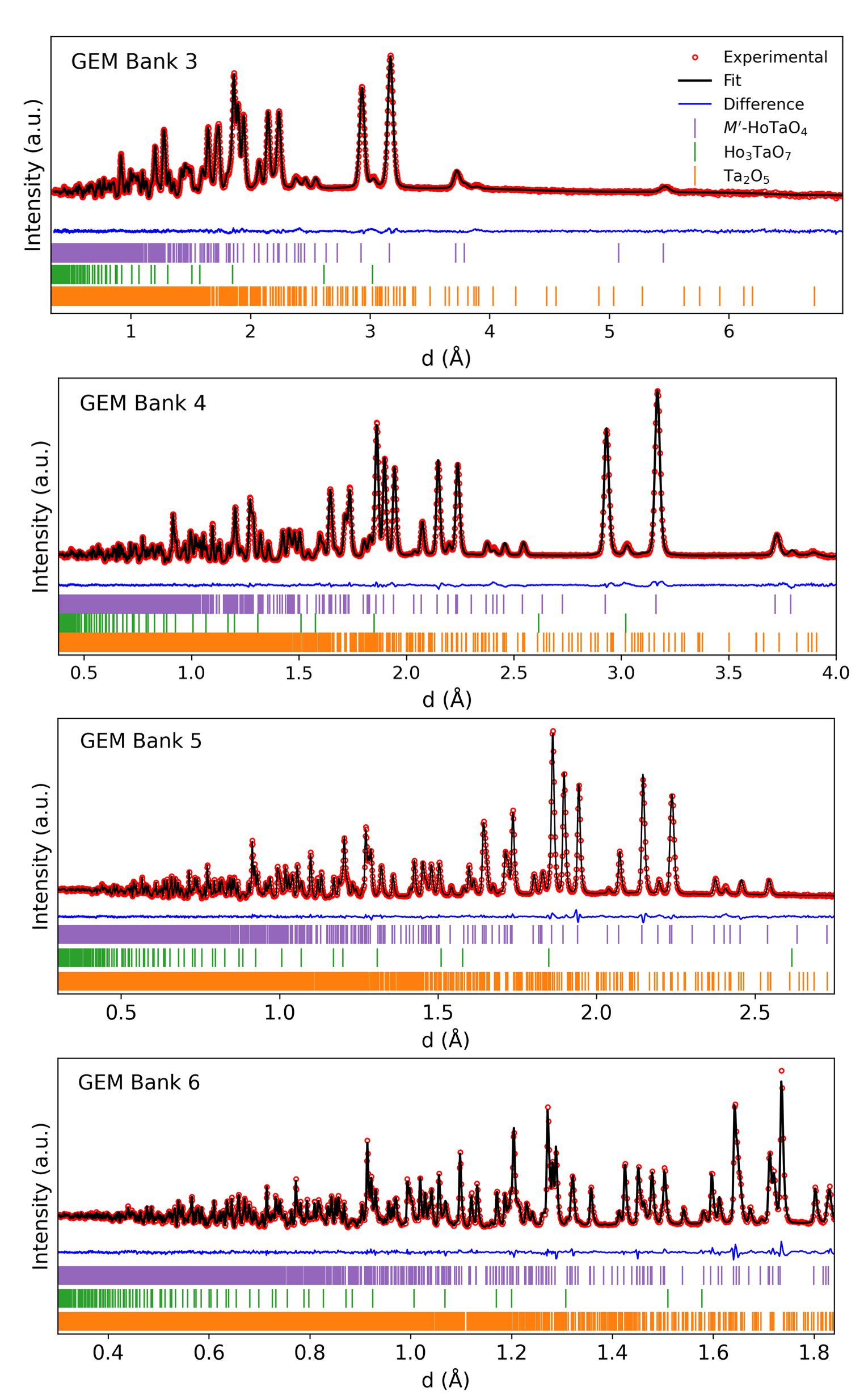}
\caption{Rietveld refinement against room-temperature TOF PND data for $M'$-\ch{HoTaO4}.}
\label{fig:rietveld_ho_pnd_rt}
\end{figure}

\begin{figure}[htbp]
\centering
\includegraphics[width=0.8\textwidth]{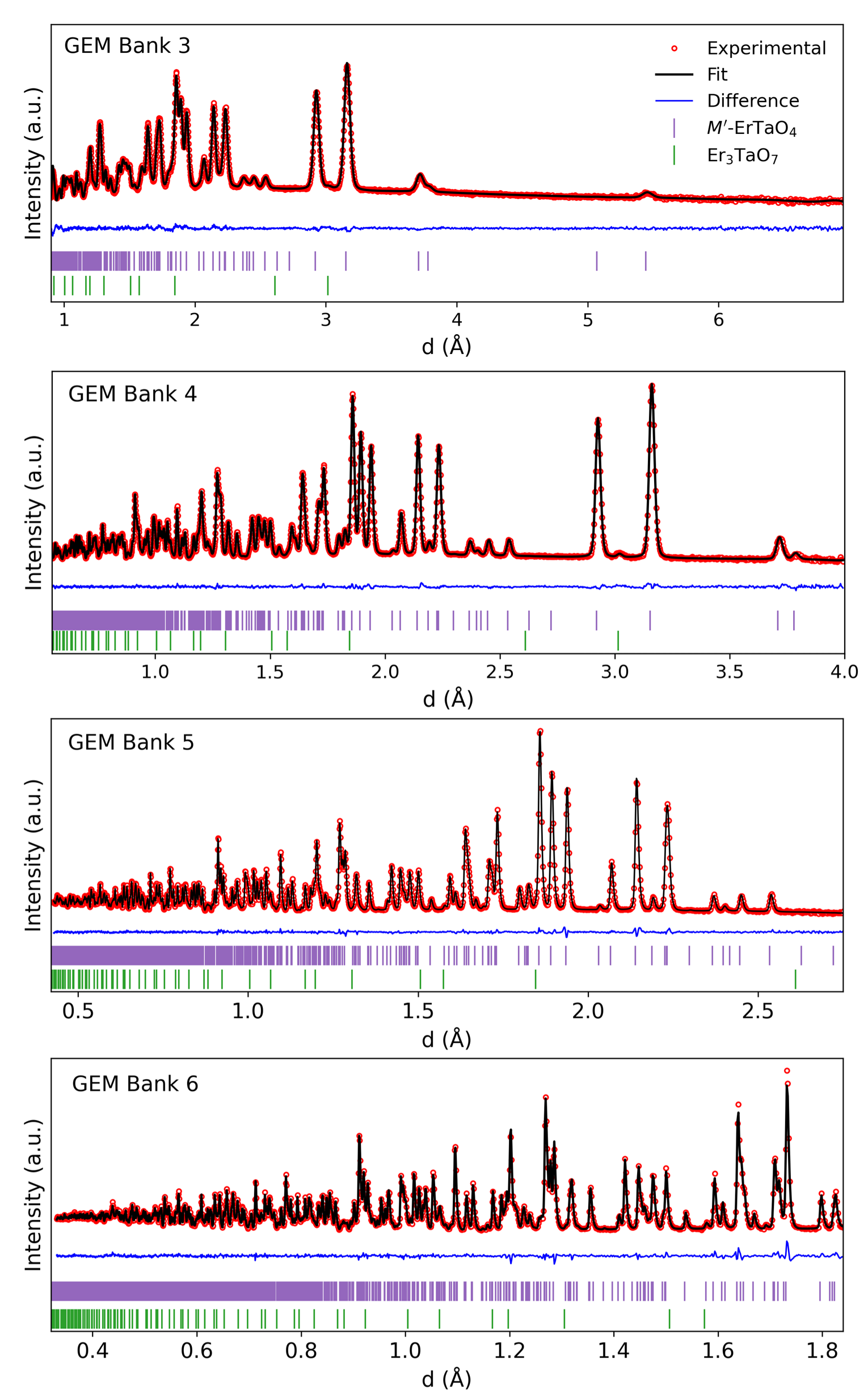}
\caption{Rietveld refinement against room-temperature TOF PND data for $M'$-\ch{ErTaO4}.}
\label{fig:rietveld_er_pnd_rt}
\end{figure}

\begin{figure}[htbp]
\centering
\includegraphics[width=0.9\textwidth]{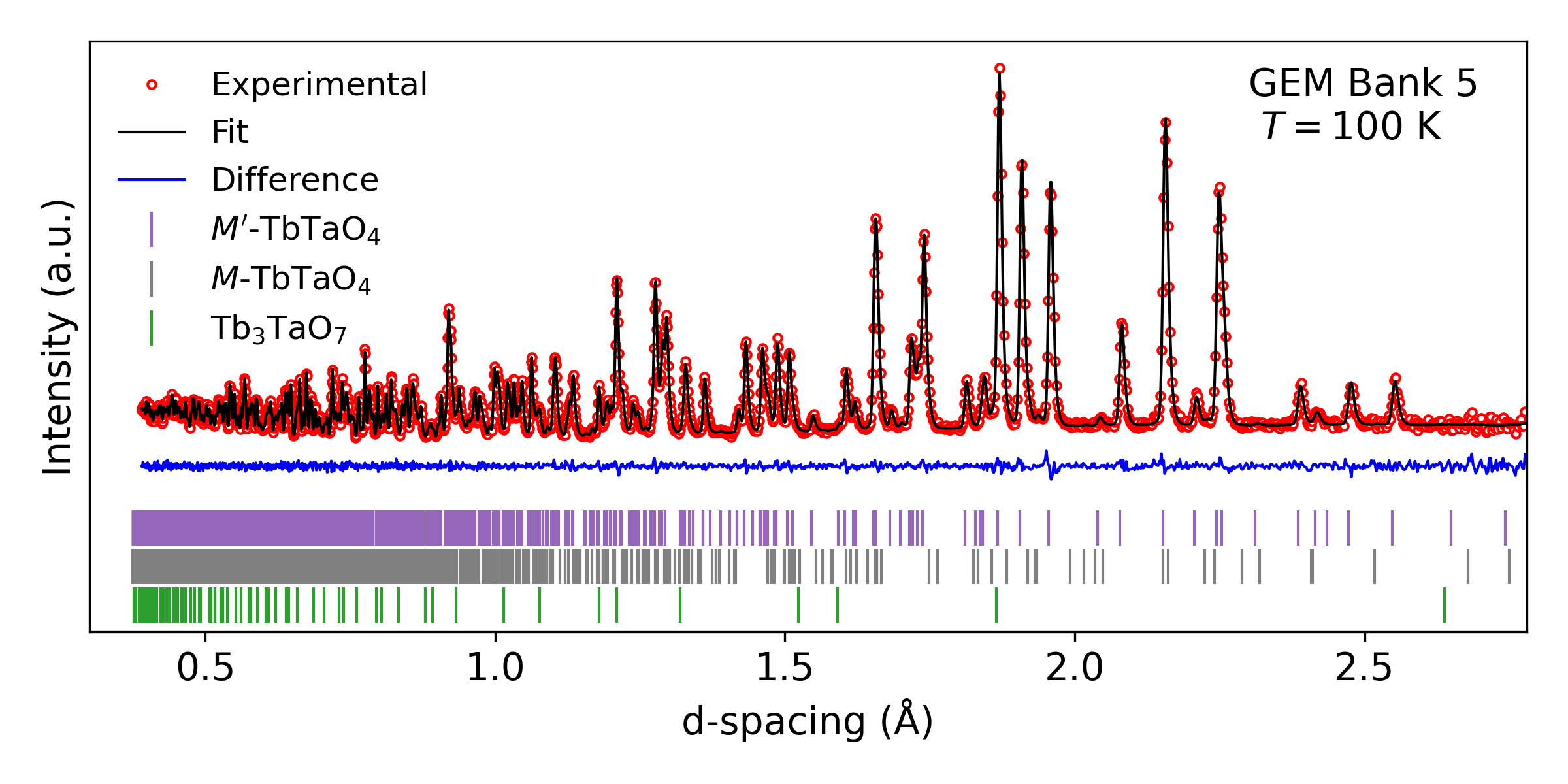}
\caption{Rietveld refinement against TOF PND data for $M'$-\ch{TbTaO4} at 100~K.}
\label{fig:100K_Mprime_TbTaO4_bank5}
\end{figure}

\begin{figure}[htbp]
\centering
\includegraphics[width=0.9\textwidth]{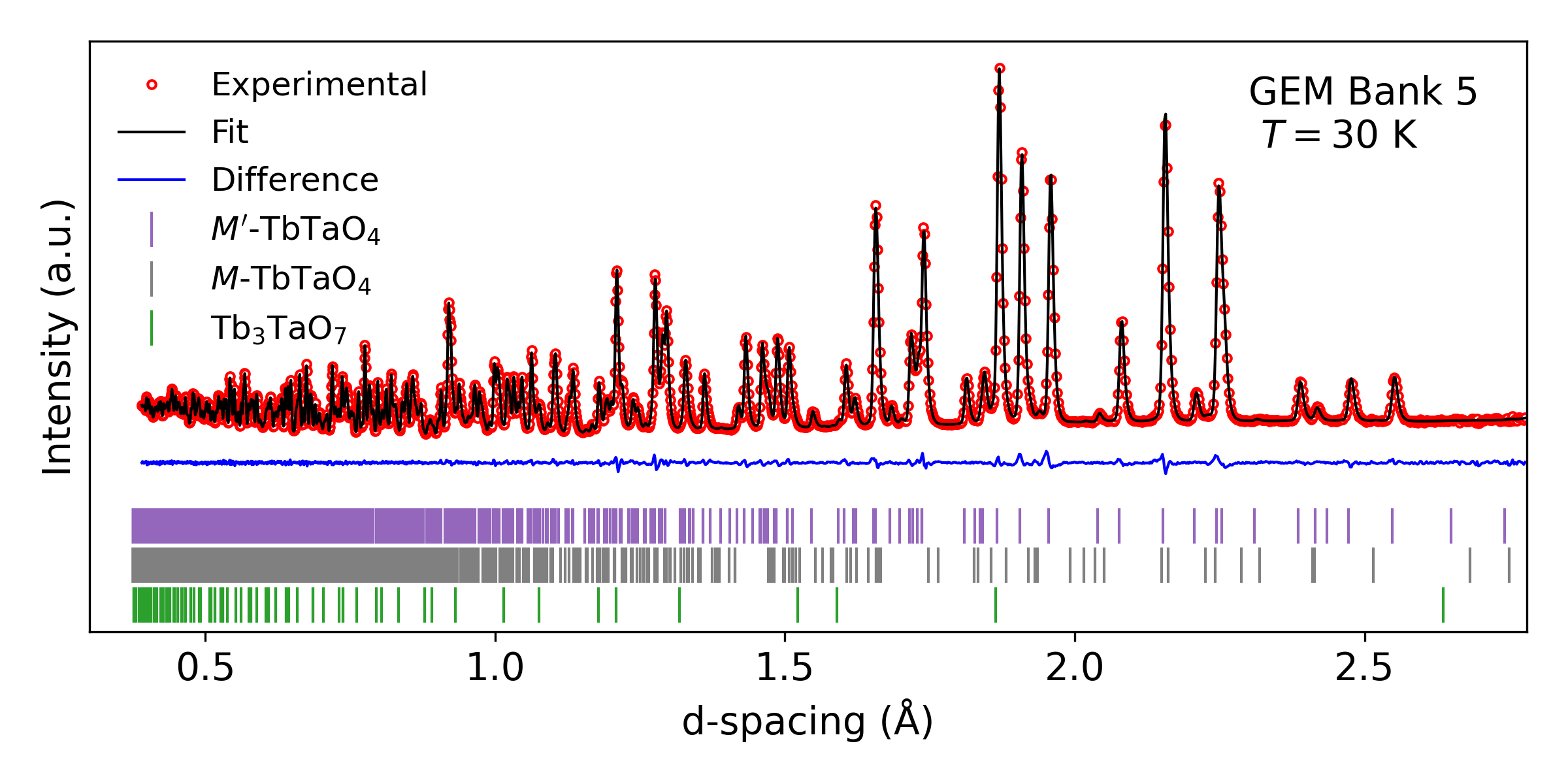}
\caption{Rietveld refinement against TOF PND data for $M'$-\ch{TbTaO4} at 30~K.}
\label{fig:30K_Mprime_TbTaO4_bank5}
\end{figure}

\begin{table}[htbp]
\centering
\caption{Refined structural data for $M'$-\ch{TbTaO4} as a function of temperature, from TOF PND data. Space group: $P2/c$. Tb atom at (0, $y_\mathrm{Tb}$, 0.25); Ta1 atom at (0.5, $y_\mathrm{Ta}$, 0.75).}
\label{table:tbtao4_pnd_vt}
\resizebox{\columnwidth}{!}{
\begin{ruledtabular}
\begin{tabular}{c c c c c c}
\multicolumn{2}{c}{$T$ (K)} & 1.5 & 30 & 100 & 300  \\
\midrule
\multicolumn{2}{c}{$a$ (\AA)} & 5.13041(3) & 5.13083(3) & 5.13202(6) & 5.14258(2) \\
\multicolumn{2}{c}{$b$ (\AA)} & 5.48437(4) & 5.48444(4) & 5.48456(7) & 5.49176(2) \\
\multicolumn{2}{c}{$c$ (\AA)} & 5.33479(3) & 5.33512(2) & 5.33484(6) & 5.33763(2) \\
\multicolumn{2}{c}{$\beta$ (\degree)} & 96.7799(4) & 96.7771(5) & 96.7594(9) & 96.6566(3) \\
\multicolumn{2}{c}{$V$ (\AA$^3$)} & 149.056(2) & 149.080(2) & 149.115(3) & 149.728(1) \\
\midrule
Tb1 & $y$ & 0.73448(4) & 0.73460(9) & 0.73436(16) & 0.73439(5) \\
Ta1 & $y$ & 0.80844(7) & 0.80833(9) & 0.80851(17) & 0.80767(6) \\
O1 & $x$ & 0.74857(6) & 0.74872(8) & 0.74889(15) & 0.74789(5) \\
   & $y$  & 0.41258(6) & 0.41269(8) & 0.41278(14) & 0.41318(5) \\
   & $z$  & 0.39744(6) & 0.39731(9) & 0.39720(16) & 0.39731(5) \\
O2 & $x$  & 0.27061(6) & 0.27056(9) & 0.27050(16) & 0.27091(6) \\
   & $y$  & 0.06471(6) & 0.06513(8) & 0.06480(15) & 0.06469(5) \\
   & $z$  & 0.49557(6) & 0.49579(9) & 0.49596(16) & 0.49543(6) \\
\midrule
\multicolumn{2}{c}{$R_\mathrm{wp}$ (\%)} & 1.98 & 2.18 & 4.40 & 1.41 \\
\multicolumn{2}{c}{$\chi^2$} & 2.33 & 4.37 & 1.48 & 3.19 \\
\end{tabular}
\end{ruledtabular}
}
\end{table}

\begin{figure}[htbp]
\centering
\includegraphics[width=0.5\textwidth]{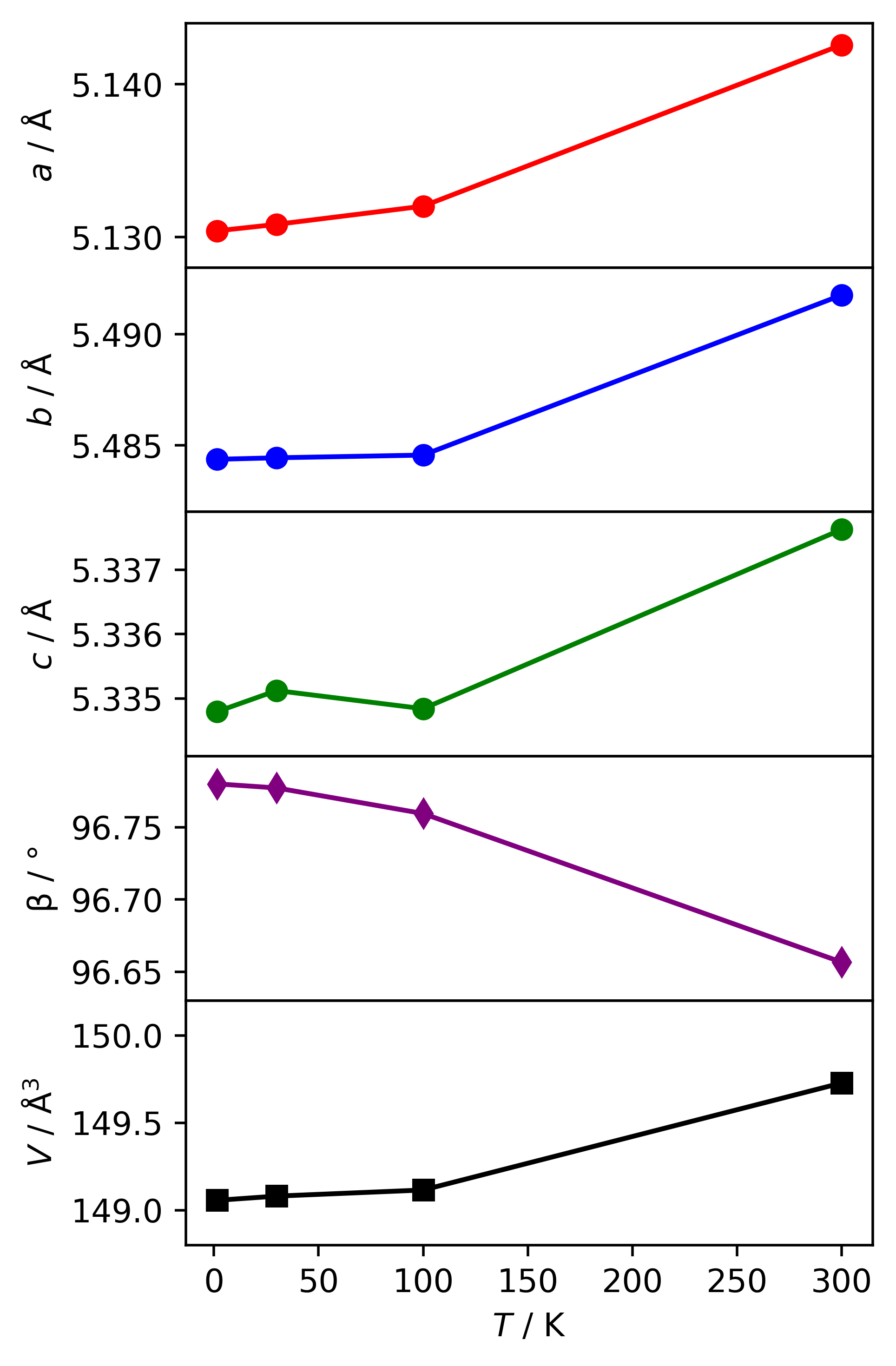}
\caption{Refined unit cell dimensions for $M'$-\ch{TbTaO4} as a function of temperature, obtained by Rietveld refinement against TOF PND data. Error bars are smaller than the datapoints.}
\label{fig:Mprime_TbTaO4_neutron_lps_VT}
\end{figure}

\begin{figure}[htbp]
\centering
\includegraphics[width=0.9\textwidth]{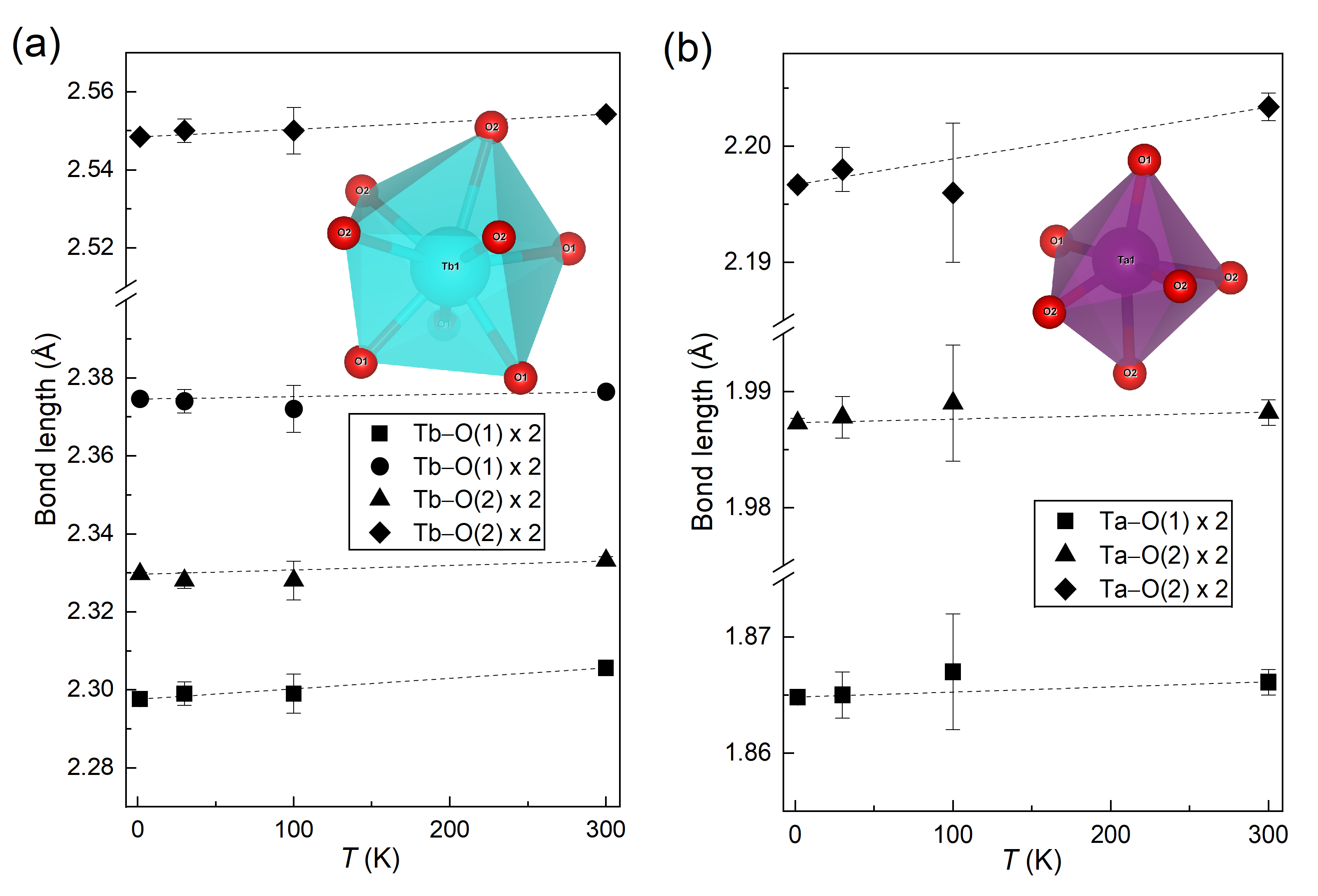}
\caption{Refined metal-oxygen bond distances for $M'$-\ch{TbTaO4} as a function of temperature, obtained by Rietveld refinement against TOF PND data. Dashed lines are the lines of best fit.}
\label{fig:Mprime_TbTaO4_neutron_bondlengths}
\end{figure}

\begin{figure}[htbp]
\centering
\includegraphics[width=0.8\textwidth]{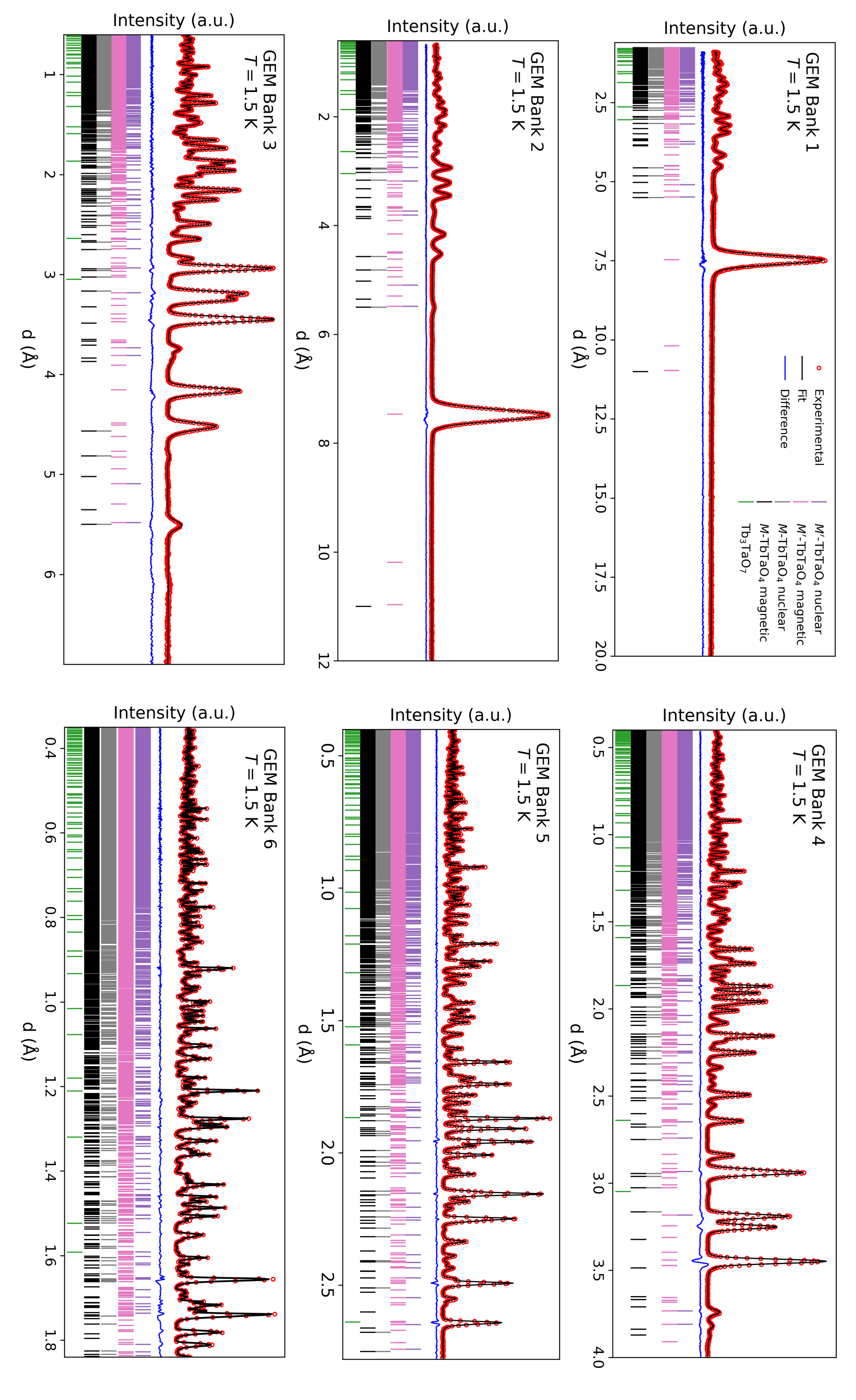}
\caption{Rietveld refinement against TOF PND data for $M'$-\ch{TbTaO4} at 1.5~K, banks 1--6.}
\label{fig:rietveld_tb_pnd_cold_allbanks}
\end{figure}

\begin{table}[htbp]
\centering
\caption{Combined nuclear and magnetic structure model for $M'$-\ch{TbTaO4} at 1.5~K, suitable for refinement in TOPAS using magnetic space group 14.80. To obtain the starting atomic positions for the combined model, the $x$ and $y$ fractional coordinates of the atoms in the nuclear unit cell were halved in order to account for the doubling of the cell in the $a$ and $b$ directions. Furthermore, two additional oxygen ions were required because of a difference in multiplicities of the Wyckoff sites. The pairs of atom sites O1/O3 and O2/O4 are each related by a translation of $(\frac{1}{2},\frac{1}{2},0)$.}
\label{table:tbtao4_mag_structure}
\resizebox{\columnwidth}{!}{
\begin{ruledtabular}
\begin{tabular}{c c c c c}
$a$ (\AA) & \multicolumn{4}{c}{10.2616(13)} \\
$b$ (\AA) & \multicolumn{4}{c}{10.9712(14)} \\
$c$ (\AA) & \multicolumn{4}{c}{5.3358(7)} \\
$\beta$ (\degree) & \multicolumn{4}{c}{96.776(2)} \\
$V$ (\AA$^3$) & \multicolumn{4}{c}{596.51(13)} \\
\midrule
 & $x$ & $y$ & $z$ & $B_\mathrm{iso}$ (\AA$^2$) \\
\midrule
Tb1 & 0 & 0.36726(8) & 0.25 & 0.017(4) \\
mlx, mly, mlz input ($\mu_\mathrm{B}$\,\AA$^{-1}$) & $-0.0683(8)$ & 0 & $1.5095(12)$ & -- \\
Crystal Axes output ($\mu_\mathrm{B}$) & $-0.702(8)$ & 0 & $8.053(6)$ & -- \\
Cartesian Axes output ($\mu_\mathrm{B}$) & $-1.652(12)$ & 0 & $7.997(6)$ & -- \\

Ta1 & 0.25 & 0.40388(15) & 0.75 & 0.081(5) \\
O1 & 0.37415(14) & 0.20587(13) & 0.3979(3) & 0.240(3) \\
O2 & 0.13517(14) & 0.03238(12) & 0.4956(3) & 0.240(3) \\
O3 & 0.87415(14) & 0.70587(13) & 0.3979(3) & 0.240(3) \\
O4 & 0.63517(14) & 0.53238(12) & 0.4956(3) & 0.240(3) \\
\midrule
$R_\mathrm{wp}$ (\%) & \multicolumn{4}{c}{1.98} \\
$\chi^2$ & \multicolumn{4}{c}{2.33} \\
\end{tabular}
\end{ruledtabular}
}
\end{table}

\clearpage

\section{\label{app:magnetic}Additional susceptibility data}

\begin{figure}[htbp]
\centering
\includegraphics[width=0.8\textwidth]{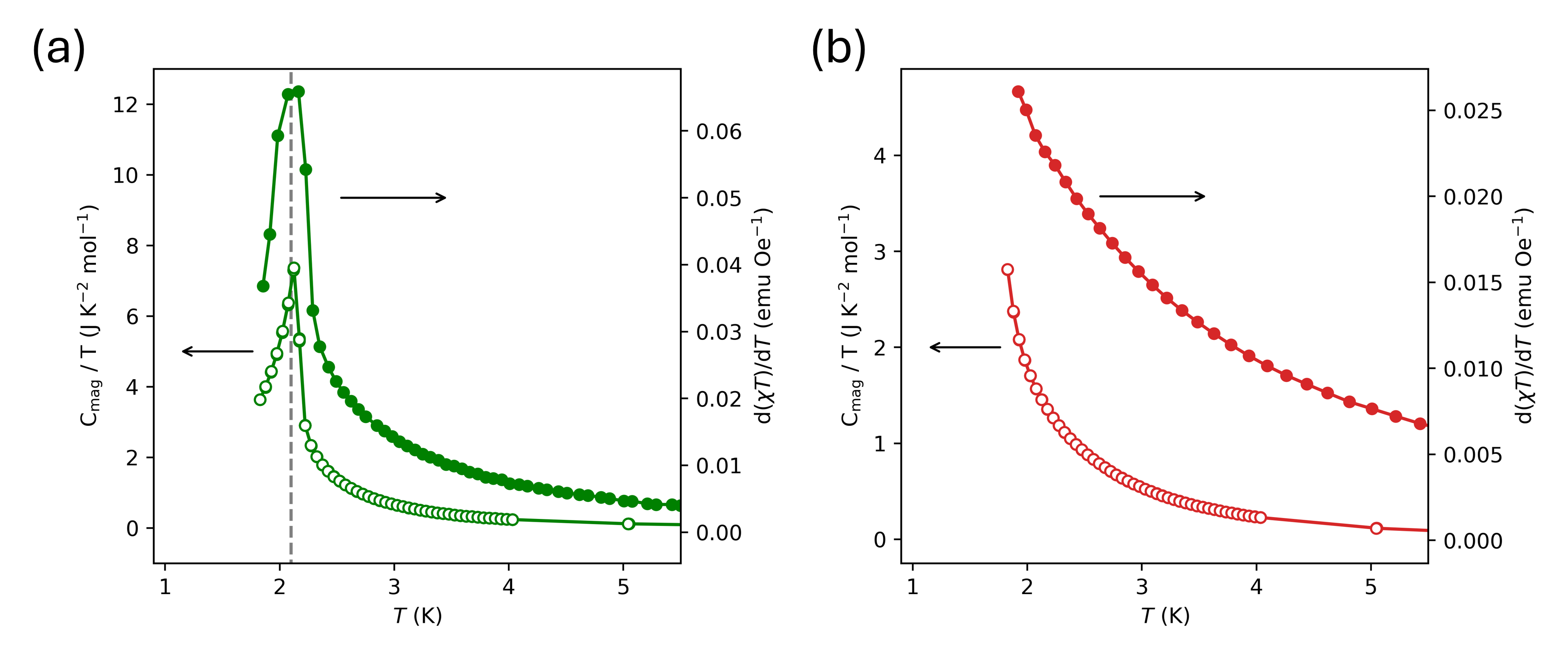}
\caption{(a) Comparison of the heat capacity data with the derivative of the susceptibility, $d(\chi T)/dT$, for $M'$-\ch{TbTaO4}. The peaks in the two plots occur at the same temperature, marked with a grey dashed line, consistent with expected behaviour for a long-range ordered antiferromagnet \cite{Fisher1962}. (b) Heat capacity data and $d(\chi T)/dT$ for $M'$-\ch{DyTaO4}. There is no peak in either the heat capacity or $d(\chi T)/dT$.}
\label{fig:ChiT}
\end{figure}

\begin{figure}[htbp]
\centering
\includegraphics[width=\textwidth]{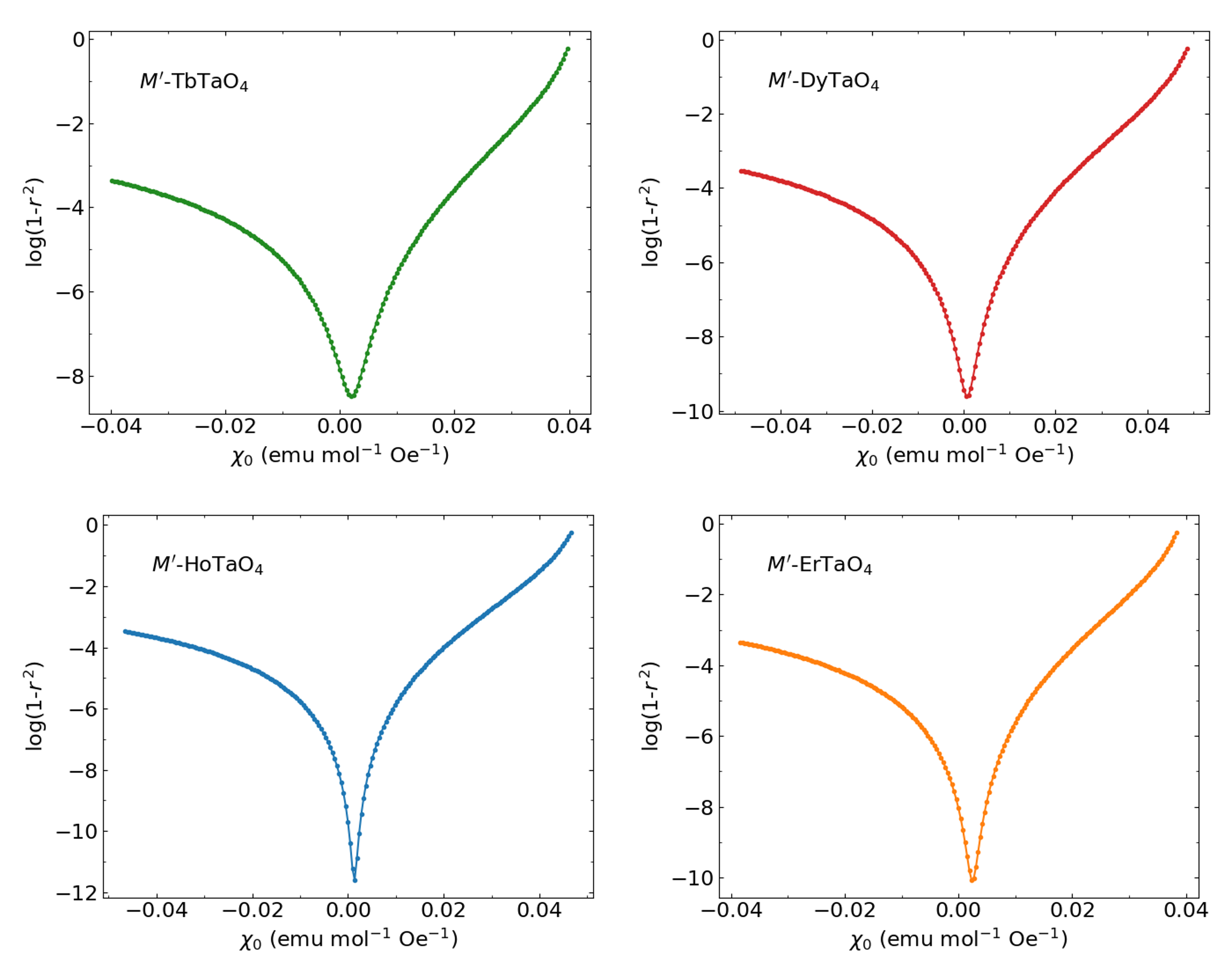}
\caption{Optimisation of the fixed susceptibility contribution $\chi_0$ for the four $M'$-\ch{\textit{Ln}TaO4} compounds in this study.}
\label{fig:nm}
\end{figure}

\end{document}